\documentclass[review]{elsarticle}

\usepackage{lineno,hyperref}
\modulolinenumbers[5]

\usepackage{graphicx}
\graphicspath{{./Images/}}

\usepackage{amssymb}

\newcommand{\fg}[1]{\mbox{\pmb{$#1$}}}
\newcommand{\bey}{\begin{eqnarray}}
\newcommand{\eey}{\end{eqnarray}}
\newcommand{\vep}{\varepsilon}
\newcommand{\fvep}{\fg \varepsilon}
\newcommand{\fsg}{\fg \sigma}
\newcommand{\sg}{\sigma}
\newcommand{\bec}{\begin{center}}
\newcommand{\eec}{\end{center}}
\newcommand{\bs}[1]{\mbox{\pmb{$#1$}}}
\newcommand{\drop}[1]{}

\usepackage{amsbsy}
\usepackage{amsmath}
\usepackage{amsfonts}
\usepackage{amssymb}
\usepackage{textcomp}
\usepackage{cleveref}
\usepackage[margin=2.5cm]{geometry}
\usepackage{xcolor}
\usepackage{rotating}
\usepackage{float}
\usepackage{subfig}

\DeclareMathOperator{\tr}{tr}

\Crefname{figure}{\text{Fig.}}{\text{Figs.}}
\Crefname{equation}{\text{Eq.}}{\text{Eqs.}}

\begin{document}

\begin{frontmatter}

        \title{Analytical and Scale-Free Phase-Field Studies of \texorpdfstring{$\alpha$}{alpha} to \texorpdfstring{$\omega$}{omega} Phase Transformation in Single Crystal Zirconium under Nonhydrostatic Loadings}

        \author{Raghunandan Pratoori\fnref{label1,label4}}		
        \author{Hamed Babaei\fnref{label1}}
        \author{Valery I.
        Levitas\corref{cor1}\fnref{label1,label2}}
		\ead{vlevitas@iastate.edu}
        \fntext[label1]{Department of Aerospace Engineering, Iowa State University, Ames, Iowa 50011, USA}
		\fntext[label2]{Department of Mechanical Engineering, Iowa State University, Ames, Iowa 50011, USA}
        \cortext[cor1]{Corresponding author}

        \begin{abstract}
        Zirconium (Zr)	 is an important engineering material with numerous practical applications.
        It undergoes martensitic $\alpha$ to $\omega$ phase transformation (PT) at pressures that vary from 0.67 GPa to 17 GPa under different loading conditions.
        Despite numerous experimental and theoretical studies, the effect of the nonhydrostatic stresses is not well understood.
        To separate the effect of nonhydrostatic stresses from the plastic deformation, a scale-free phase field approach (PFA) for multivariant $\alpha$ to $\omega$ PT in a single crystal Zr under general nonhydrostatic loadings is presented.
        Explicit conditions for the direct and reverse PTs between austenite and martensitic variants and between martensitic variants under general stress tensor are derived and analyzed.
        In particular, the effect of the deviatoric stresses on the PT pressures is elucidated.
        It is shown that their effect cannot explain much larger reduction in the  transformation pressure observed during plastic flow,  i.e., specific mechanisms of strain-induced phase transformations should be involved.
        Under assumption of the homogeneous fields in the sample,  complete analytical solutions that include stress-strain curves during the PT, PT start and finish stresses (i.e., stress hysteresis), and volume fraction of the variants, are determined for different loadings.
        Finite element method (FEM) solutions are found for the phase field simulations of the microstructure evolution for the same loadings, as well as for two grains of the polycrystalline sample.
        Macroscopic averaged characteristics of the PFA solutions are well described by an analytical solutions, which also simplifies their interpretations.
        Obtained results are in good qualitative agreement with existing experiments.
        In addition, some controversies of the previous approaches are analysed.
    \end{abstract}

\end{frontmatter}

    \section{Introduction}
         In the quest to enhance the reliability and performance of zirconium (Zr) and its alloys, pivotal materials in nuclear reactors and the chemical industry, understanding the microstructural evolution and phase stability is paramount.
         These materials are subjected to various thermomechanical processes to achieve optimal mechanical properties and microstructure.
         Under high hydrostatic pressure and nonhydrostatic loadings,  Zr undergoes the well-known martensitic  hcp ($\alpha$) to simple hexagonal ($\omega$) phase transformation (PT), which is typical  for group IV transition elements (including Ti  and Hf) and their alloys~\cite{Sikka-Vohra-PMS-82}.
         Due to its fundamental and practical importance, this PT is broadly studied in literature under hydrostatic  ~\cite{velisavljevic2011effects,popov2019real,Levitasetal-grainGrowth-24,Anzellini-etal-20,Pandey-Levitasetal-JAP-Zr-24}, nonhydrostatic (or quasi-hydrostatic) ~\cite{Liu-etal-23,Kumaretal-Acta-20,Kumaretal-Scripta-24},  and dynamic loadings~\cite{Swinburneetal-PRB-16,Grivickasetal-JAP-22,Greeffetal-PRB-22}  as well as during severe plastic deformations~\cite{Zilbershtein-75,Blank2013,Edalati-etal-Zr-09,Srinivasarao-etal-Zr-11,Pandey-Levitas-2020,Levitas-etal-NatCom-23,Dhar-Levitas-NPJ-kinetics-24,Lin-Levitas-ResSquare-22}.
         PT in Zr may occur  during fabrication and service,  significantly impacting mechanical properties.
         Thus, $\alpha$ Zr is weaker but more ductile, while $\omega$ Zr is stronger but brittle.

        Under hydrostatic loading in diamond anvil cell (DAC), this phase transformation occurs in the pressure range of 5 - 17 GPa ~\cite{velisavljevic2011effects,popov2019real,Levitasetal-grainGrowth-24,Anzellini-etal-20,Pandey-Levitasetal-JAP-Zr-24}.
        Such large  discrepancy is caused by different impurities and pressure-transmitting medium, which causes different degrees of non-hydrostaticity.
        Nonhydrostatic stresses significantly reduce the PT pressure in Zr, as summarized in ~\cite{Liu-etal-23}.
        The same trend is observed for the effect of nonhydrostatic stresses on $\alpha-\omega$ PT in Ti~\cite{Errandonea-etal-Ti-05}.
        The problem is that nonhydrostatic stress state in DAC is quantitatively not characterized, and degree of non-hydrostaticity is qualitatively characterized by the referring to  the strength of the  pressure-transmitting medium.

        Under plastic straining,   $\alpha-\omega$ PT in  Zr occurs in the pressure range of  0.67 - 7 GPa~\cite{Zilbershtein-75,Blank2013,Edalati-etal-Zr-09,Srinivasarao-etal-Zr-11,Pandey-Levitas-2020,Levitas-etal-NatCom-23,Dhar-Levitas-NPJ-kinetics-24,Lin-Levitas-ResSquare-22} which shows largely scattered data from various researchers; some results are even contradictory.
        These results have been obtained mostly by high-pressure torsion with ceramic, metallic, or diamond anvils.
        However, in~\cite{Pandey-Levitas-2020}, the same PT pressure for strongly pre-deformed by rolling Zr was obtained by high-pressure torsion in rotational DAC (RDAC) and compression without pressure-transmitting medium.
        That is why most of DAC experiments   without pressure-transmitting medium involve significant plastic deformation, and reduction in PT pressure is not only due to nonhydrostatic stresses, but mostly due to plastic straining.
        Plastic strain-induced PTs under high pressure require completely different experimental characterization and thermodynamic and kinetic treatment than the pressure- and stress-induced PTs ~\cite{Levitas-PRB-2004,levitas2018high,Levitas_2019,Levitas-IJP-21,Levitas-MT-23}.

        Thus, the effect of nonhydrostatic stresses on  $\alpha-\omega$ PT in  Zr  (and other materials) is not quantitatively characterized, and in many cases it is unclear whether the effect is due to nonhydrostatic stresses, or plastic straining, or both.
        Thermodynamics of non-hydrostatically loaded Zr was developed in~\cite{Zhangetal-SciRep-19} with analysis of the effect on nonhydrostatic (deviatoric) stresses on the phase equilibrium lines between $\alpha$,  $\omega$, and $\beta$ phases.
        However, the main contributor to the effect of deviatoric stresses, deviatoric transformation strain, was not included.
        Phase-field approach (PFA) study of the microstructure evolution using during $\alpha-\omega$ PT in  Zr is limited to the hydrostatic loading ~\cite{yeddu2016alpha,Yeddu-Zr-22}, which still causes internal nonhydrostatic stresses and plastic deformations due to transformation strain.

         Recognizing the criticality of the problem, we utilize here  a finite-strain, scale-free PFA to study the effect of different non-hydrostatic loadings on the $\alpha-\omega$ PT and microstructure evolution in  a single crystal Zr.
         This approach includes the crystallography of the mutlivariant PT and effect on the stress tensor on the thermodynamic driving force for PT and, consequently, on the PT thermodynamics, kinetics, and   microstructure evolution.
         Advantage of the scale-free approach is not only that it allows to study sample of an arbitrary size, in contrast to nanoscale sample in the traditional PFA~\cite{yeddu2016alpha,Yeddu-Zr-22}.
         From the macroscopic phase transformation  conditions that the thermodynamic driving force for PT is equal to the athermal dissipative threshold and assumption of the homogeneous fields in the sample, we also  can find complete analytical solution  including stress-strain curves during the PT, PT start and finish stresses (i.e., stress hysteresis), and volume fraction  of the variants.
         Because of including a barrier term in the expression for the free energy and athermal dissipative threshold, PFA treatment is more general than the traditional thermodynamic approach~\cite{Zhangetal-SciRep-19} which considers equilibrium at fixed stresses.
         Analytical solution describes well overall behavior of the sample from the  results of the PFA simulations even when microstructure is heterogeneous.
         Independently, obtained expressions for the thermodynamic driving forces for austenite - martensitic variant $i$ ($i=1,\,2,\,3$) PT and transformations between martensitic variants under general stress tensor allowed us to analytically obtain and analyse the effect of the stress deviator on these transformations.


        Here, we explore the application of a scale-free PFA for the simulation of discrete martensitic microstructures.
        Initially, a scale-free PFA was developed for small strains and cubic to tetragonal PT by~\cite{Levitas-etal-2004,Idesman-etal-2005}, and later was refined and utilized for cubic to monoclinic PT in NiTi shape memory alloys by~\cite{Esfahani-etal-2018}.
        The finite-strain extension of the scale-free model has been introduced by~\cite{Levitas-IJP-18,Babaei-Levitas-IJP-18,Babaei-Levitas-JMPS-20,babaei2023simulations}.
        This framework has been employed in simulating  cubic to tetragonal PT in single and polycrystalline silicon from Si I to Si II.

        The primary distinctions between the scale-free model and nanoscale PFAs encompass:
        \begin{enumerate}
            \item Omission of the gradient energy term related to order parameter gradients, which defines phase interface widths and energies, rendering the model scale-free.
            Consequently, the interface width matches a single finite element, enhancing computational efficiency compared to the nanoscale model that typically requires 3-5 elements to mimic an analytical interface solution.
            However, this simplification introduces two challenges: (a) Solution dependency on the mesh size; however, extensive computational studies by~\cite{Levitas-etal-2004,babaei2020finite} have demonstrated that the solution stabilizes when the mesh is reduced to 1/80th of the sample size, becoming nearly mesh-independent. (b) The presence of significant strain gradients within the one-element thick interface due to the variation of martensite volume fraction from 0 to 1, often causing divergence in the finite element method (FEM) solution.
            \item Non-resolution of interfaces between different martensitic variants, given that each variant's width is approximately 10 nm, translating to thousands of interfaces at the microscale and rendering the computational task unfeasible.
            Instead, martensite is treated as a composite of variants in proportion to their volume fractions.
            \item The application of a linear mixture rule for material properties instead of complex polynomials used in nanoscale PFA, as the atomic-level energy landscape against order parameters is not replicated.
            This approach simplifies the inclusion of an athermal dissipative threshold for interface movement (akin to interface friction), a challenge in nanoscale models.
            \item Utilization of the martensite volume fraction {$c$ }as the order parameter, resulting in material instability and transformation strain localization producing  the austenite-martensite interfaces.
            The volume fractions of individual martensitic variants {$c_i$ } act merely as internal variables without inducing instabilities, thereby removing the necessity for interfaces between variants.
        \end{enumerate}
        These strategies collectively facilitate the efficient simulation of multivariant martensitic phase transformations in samples of any size.

        In this study, we expand upon the modeling techniques introduced by~\cite{babaei2020finite} to simulate the $\alpha$ to $\omega$ PT in single-crystal Zr.
        A significant hurdle we address is achieving computational convergence, challenged by the shear transformational strains observed in the second and third variants and large strain gradients.
        Our exploration includes a variety of loading conditions, ranging from uniaxial to hydrostatic pressures, differing in the number of elements employed, as well as examining complex loading scenarios exemplified by an isolated grain within a polycrystalline matrix.
        Mesh-dependence of the FEM solutions is analysed in detail.
        Analytical solutions are verified by FEM simulations and also help interpret FEM results.
        Obtained results are in good qualitative agreement with existing experiments.
        Also, some controversies of the previous approaches are analysed.

    \section{Model description}\label{model}
        Here, we summarize  the microscale model for multivariant martensitic PTs developed in~\cite{babaei2020finite}.
        Vectors and tensors are designated with boldface symbols.
        The contractions of tensors ${\fg A}=\{A^{ij}\}$ and ${\fg B}=\{B^{ji}\}$ over one and two indices is designated as ${\fg A}{\fg \cdot}{\fg B}=\{A^{ij}\,B^{jk}\}$ and ${\fg A}{\fg :}{\fg B}=A^{ij}\,B^{ji}$.
        The transpose,  inverse, and symmetric part of ${\fg A}$ are ${\fg A}^{ T}$, ${\fg A}^{-1}$, and $({\fg A})_s$, respectively; ${\fg I}$ is the unit tensor; $\fg \nabla_0$ and $\fg \nabla$ are the gradient operators in the undeformed and deformed states, respectively;  $\fg \nabla_0 \bs{\cdot} $ and $\fg \nabla \bs{\cdot}$ are the divergence operators in the undeformed and deformed states, respectively.

        \subsection{Kinematics}

            We start with a single crystal of zirconium ($\alpha$ phase) in its stress-free configuration, denoted as $\Omega_0$, described by the position vector $\mathbf{r}_0$.
            The deformation of the crystal, including phase transformation to the $\omega$ phase, can be described by a continuous function $\mathbf{r}(\mathbf{r}_0, t)$, where
            $\mathbf{r}$ represents the position vector in the current deformed configuration $\Omega$ and $t$ is the time.
            We define  the deformation gradient $\mathbf{F}$ and its  multiplicative decomposition into two components, the elastic part denoted as $\mathbf{F}_e$ and the symmetric transformational part denoted as $\mathbf{U}_t$:
            \begin{equation}
                {\bs F} := \bs \nabla_0 {\bs r}= {\bs F}_e{\bs{\cdot}}{\bs U}_t;    \qquad {\bs F}_e={\bs R}_e\cdot{\bs U}_e.
                \label{Fdecom}
            \end{equation}
            Here, $\mathbf{U}_e$ represents the symmetric elastic right stretch tensor, and $\mathbf{R}_e$ is the orthogonal lattice rotation tensor that occurs during loading and phase transformation.			In this context, tensors  $\mathbf{U}_t$ generate the local intermediate stress-free configuration $\Omega_t$ and are  defined 
            as
            \begin{equation}
                \bs{U}_t= \bs{I}+ \fg{\varepsilon}_{t}=
                \bs{I}+\sum_{i=1}^m {\fg{\varepsilon}}_{ti} c_i.
                \label{tr-str}
            \end{equation}
             Here, $\fg{\varepsilon}_t$ represents the transformation strain,  $\boldsymbol{\varepsilon}_{ti}$ is the transformation strain of the $i^{th}$ martensitic variant, and $m$ is the number of variants.
             The symbol $c_i$ corresponds to the volume fraction of the $i^{th}$ variant expressed via  volumes in $\Omega_0$.
            Defining the total, elastic, and transformational Lagrangian strains as
            \begin{equation}
                \bs{E}=\frac{1}{2}(\bs{F}^T{\bs{\cdot}}\bs{F}-\bs{I});\quad
                \bs{E}_e=\frac{1}{2}(\bs{F}_e^T{\bs{\cdot}}\bs{F}_e-\bs{I});\quad
                \bs{E}_t=\frac{1}{2}(\bs{F}_t^T{\bs{\cdot}}\bs{F}_t-\bs{I}), \qquad
                \label{Esall}
            \end{equation}
            we can connect them as
            \begin{equation}
                \bs{E}_e = \bs{F}_t^{-1}{\bs{\cdot}}		(\bs{E}-\bs{E}_t){\bs{\cdot}} \bs{F}_t^{-1}
                \label{Erel}
            \end{equation}
            by utilizing the multiplicative decomposition given in \Cref{Fdecom}.				
            Since strains in our simulations do not exceed 0.1, small strain approximation may deliver acceptable results from engineering point of view accuracy, while significantly extending a pool of material scientists who can access the equations without knowledge of nonlinear continuum mechanics.
            For small strains, instead of \Cref{Fdecom,Erel}, we obtain:
            \begin{equation}\label{e26}
                \bs{\varepsilon}:=\left(  \bs \nabla {\bs u}\right)_s=\bs{\varepsilon}_e+\bs{\varepsilon}_t;
                \qquad
                \bs{\varepsilon}_{t}=
                \sum_{i=1}^m {\bs{\varepsilon}}_{ti} c_i,
            \end{equation}
            where ${\bs u}={\bs r}-{\bs r}_0$	is the displacement vector, and $	\bs{\varepsilon}$, $\bs{\varepsilon}_e$, and $\bs{\varepsilon}_t$ are small total, elastic, and transformational strains. { Trace of \Cref{e26} gives the additive relationship between volumetric total and elastic and transformation strains:
            \begin{equation}\label{e26a}
                {\varepsilon}^v={\varepsilon}_{e}^v+{\varepsilon}_{t}^v;
                \qquad
                {\varepsilon}_{t}^v= \varepsilon_{tM}^v c ; \qquad c = \sum_{i=1}^{m}c_i,
            \end{equation}
            where $ \varepsilon_{tM}^v$ is the volumetric transformation strain of any martensitic variant $M$ because they are the same. 
            \Cref{e26a} will be used for hydrostatic loading.}

        \subsection{Helmholtz free energy}
            When defining the Helmholtz free energy $\psi$ per unit undeformed volume in $\Omega_0$ for a mixture of austenite and $m$ martensitic variants, the contributions from elastic energy $\psi^{e}$, thermal energy $\psi_i^{\theta}$, and interaction energy $\psi^{in}$ are given by the following expression:
            \begin{equation} \label{energy}
                \psi(\bs{F}_e,c_i,\theta)= J_t \psi^{e}(\bs{F}_e,c_i)+\psi^{\theta}(\theta,c_i)+\psi^{in}(c_i);
            \end{equation}
            \begin{equation} \label{int_energy}
                \psi^{in} = \mathcal{A} c c_0 \geq 0;	\qquad  \qquad c_0 = 1-c;
            \end{equation}
            \begin{equation}\label{elastic_energy}
                \psi^e=\frac{1}{2}\fg {E}_e \fg :{\fg C} \fg : \fg {E}_e =\frac{1}{2}{C}^{ijkl} {E}_e^{ij} {E}_e^{kl}=\frac{1}{2}{C}^{ij} {E}_e^{i} {E}_e^{j};
            \end{equation}			
            \begin{equation}\label{thermal_energy}
                \psi^\theta=\sum_{i=0}^{m}c_i\psi_i^{\theta}(\theta)=c_0\psi_A^\theta(\theta)+c\psi_M^\theta(\theta).
            \end{equation}
            Here, $\psi^{in} = \mathcal{A} c c_0 \geq 0$ accounts for the interactions between austenite and martensite, the energy of internal stresses, as well as the energy of the austenite-martensite phase interfaces;  ${\fg C}$ is the tensor of elastic moduli of the austenite-martensite mixture.
            We will use both matrix representation of strains and stresses, and vector representation according to the Voigt notations, and corresponding representations for components of elastic moduli.
            The  interactions between different martensitic variants is omitted to avoid the formation of variant-variant interfaces.
            As we will see, a positive value of $\mathcal{A}$ leads to an increase in the thermodynamic driving force for the austenite-martensite phase transformation with increasing volume fraction of the martensite $c$ and, for the most loadings, to the  negative tangent modulus of the equilibrium stress-strain curve during the phase transformation, resulting in local mechanical instability and the formation of localized transformation bands or regions of the product phase.
            This approach reproduces a discrete martensitic structure, similar to the nanoscale PFA.

            The elastic energy
            $\psi^{e}$ is defined in $\Omega_t$ (since elasticity rule is determined experimentally in the unloaded configuration $\Omega_t$), and the Jacobian $J_t=\frac{d V_t}{dV_0}=\det{\bs F_t}$ maps it back into $\Omega_0$.
            The fourth-order tensor of elastic moduli  $\mathbf{C}$ and the  thermal (chemical) energy $\psi^{\theta}$ are determined by the mixture rule.
            We took into account that the thermal energy of  all martensitic variants is the same.
            In  \Cref{thermal_energy} $\theta$ is the temperature.
            For small strains, \Cref{energy,elastic_energy} are slightly simplified to
            \begin{equation} \label{energy-sm}
                \psi(\fvep_e,c_i,\theta)=  \psi^{e}(\fvep_e,c_i)+\psi^{\theta}(\theta,c_i)+\psi^{in}(c_i);
                \qquad
                \psi^e=\frac{1}{2}\fvep_e \fg :{\fg C} \fg :\fvep_e =\frac{1}{2}{C}^{ijkl} \vep_e^{ij} \vep_e^{kl}.
            \end{equation}	

        \subsection{Thermodynamic driving forces}
            Usual thermodynamic machinery based on the second law of thermodynamics for isothermal processes allows us to present the dissipation rate per unit undeformed volume $D$ as the sum of the product of thermodynamic driving forces $X_{i0}$ for $A \rightarrow M_i$ transformation and $X_{ij}$ transformations between martensitic variants, and their respective transformation  rates $\dot{c}_{i0}$ and $\dot{c}_{ij}$:
            \begin{align}
                &D=\sum_{i=1}^{m}X_{i0}\dot{c}_{i0}+\sum_{j=1}^{m-1}\sum_{i=j+1}^{m}X_{ij}\dot{c}_{ij} \geq 0;\\ 
                &X_{i0}=W_{i0}-\frac{J_t}{2}\bs{E}_e\bs{:}({\bs{C}}_i-{\bs{C}}_0)\bs{:}\bs{E}_e-
                \frac{J_t}{2}\left(\bs{E}_e\bs{:}{\bs{C}}(c)\bs{:}\bs{E}_e \right) {\fg F}_{t}^{-1} \bs{:} \bs{\varepsilon}_{ti}
                -( \psi^{\theta}_M-\psi^{\theta}_A) -\mathcal{A}(1-2c); 
                \\
                &X_{ij}=W_{ij}-\frac{J_t}{2}\bs{E}_e\bs{:}({\bs{C}}_i-{\bs{C}}_j)\bs{:}\bs{E}_e
                -\frac{J_t}{2}\left( \bs{E}_e\bs{:}{\bs{C}}(c)\bs{:}\bs{E}_e \right) {\fg F}_{t}^{-1} \bs{:} (\bs{\varepsilon}_{ti}-\bs{\varepsilon}_{tj});
                \\
                &W_{i0} =\bs{P}^T\bs{\cdot}\bs{F}_e\bs{:}\bs{\varepsilon}_{ti}=
                J{\fg F}^T_e {\fg \cdot} \fsg
                \bs{\cdot} {\fg F}_{e}^{t-1}  \cdot {\fg F}_{t}^{-1} \bs{:}\bs{\varepsilon}_{ti};
                \quad
                W_{ij}= \bs{P}^T\bs{\cdot}\bs{F}_e\bs{:}(\bs{\varepsilon}_{ti}-\bs{\varepsilon}_{tj})=
                J {\fg F}^T_e {\fg \cdot} \fsg
                \bs{\cdot} {\fg F}_{e}^{t-1}  \cdot {\fg F}_{t}^{-1}  \bs{:}(\bs{\varepsilon}_{ti}-\bs{\varepsilon}_{tj}).
                \label{dissipation}
            \end{align}
            In this context, $\dot{c}_{i0}$ and $\dot{c}_{ij}$ represent the rate of change of the volume fraction of variant $i$ due to the transformation {from} austenite and variant $j$, respectively; $W_{i0}$ stands for the transformation work for the austenite to martensite phase transformation, and $W_{ij}$ is the transformation work from variant $j$ to variant $i$; both are expressed in terms of
            the nonsymmetric first Piola-Kirchhoff stress
            $ \bs{P}$ and the symmetric Cauchy (true) stress tensor $\fsg=J^{-1}  \bs{P} \bs{\cdot}\bs{F}^T $; $  J=\frac{d V}{dV_0}=\det{\bs F}$.

            Simulations show that the terms related to the change in elastic moduli and $J_t$ can be neglected with acceptable error, which leads to simplified expressions for
            $X_{i0}$ and $X_{ij}$:
            \begin{align}\label{Xs}
                X_{i0}=W_{i0}
                -( \psi^{\theta}_M-\psi^{\theta}_A) -\mathcal{A}(1-2c); 
                \qquad
                X_{ij}=W_{ij}.
            \end{align}
            For small strains expressions for
            $W_{i0}$ and $W_{ij}$ are simplified to
            \begin{align}\label{Ws-sm}
                W_{i0} = \fsg \bs{\cdot}  \bs{:}\bs{\varepsilon}_{ti};
                \qquad
                W_{ij}= \fsg  \bs{:}(\bs{\varepsilon}_{ti}-\bs{\varepsilon}_{tj}).
            \end{align}

        \subsection{Transformation kinetics}
            The kinetic equations represents the linear relationships between the transformation rates and difference between the corresponding driving force and the athermal thresholds $k_{i-0}$  and $k_{i-j}$.
            We obtain  for the $ A\leftrightarrow M_i$ PTs
            \begin{equation}\label{kinetic_i0}
                \begin{cases}
                    \begin{aligned}
                        \dot{c}_{i0}=\lambda_{i0}(X_{i0}-k_{i-0}) \quad \text{if}\;
                        &\{X_{i0}-k_{i-0}>0 \;\&\; c_i<1 \;\&\; c_0>0\} \; \qquad A \rightarrow M_i\\
                        \dot{c}_{i0}=\lambda_{i0}(X_{i0}+k_{0-i}) \quad
                        \text{if}\; &\{X_{i0}+k_{0-i}<0 \;\&\; c_i>0 \;\&\; c_0<1\}\qquad \; M_i \rightarrow A
                    \end{aligned}\\
                    \dot{c}_{i0}=0 \qquad \qquad \qquad \; \text{otherwise;} \qquad i=1,2,...,m,
                \end{cases}
            \end{equation}
            and for $ M_j\leftrightarrow M_i $ PTs
            \begin{equation}\label{kinetic_ij}
                \begin{cases}
                    \begin{aligned}
                        \dot{c}_{ij}=\lambda_{ij}sign(X_{ij})(|X_{ij}|-k_{ij}) \quad \text{if}\;
                        &\{|X_{ij}|-k_{ij}>0 \;\&\; c_i<1 \;\&\; c_j>0\} \; \qquad M_i \rightarrow M_j\\
                        \text{or}\; &\{|X_{ij}|-k_{ij}<0 \;\&\; c_i>0 \;\&\; c_0<1\}\qquad \; M_i \rightarrow A
                    \end{aligned}\\
                    \dot{c}_{ij}=0 \qquad \qquad \qquad \; \text{otherwise;} \qquad i=1,2,...,m,
                \end{cases}
            \end{equation}
            where  $ \lambda_{i0} $ and $ \lambda_{ij} $ are the proportionality factors.
            To impose that the PT from any phase does not occur if this phase does not exist, we utilize the non-strict inequalities for the volume fraction of phases.
            Traditionally, $k_{i-0}=k_{0-i}$, since usually the phase equilibrium pressure is assumed to be a semisum of the pressures for the direct and reverse PTs; also,  $k_{i-0}=k_{M-A}$ for all $i$.
            When the athermal thresholds for PTs are neglected, kinetic equations simplify to
            \begin{equation}\label{kinetic_i0-2}
                \begin{cases}
                    \begin{aligned}
                        \dot{c}_{i0}=\lambda_{i0}X_{i0} \quad \text{if}\;
                        &\{X_{i0}>0 \;\&\; c_i<1 \;\&\; c_0>0\} \; \qquad A \rightarrow M_i\\
                        \text{or}\; &\{X_{i0}<0 \;\&\; c_i>0 \;\&\; c_0<1\}\qquad \; M_i \rightarrow A
                    \end{aligned}\\
                    \dot{c}_{i0}=0 \qquad \qquad \qquad \; \text{otherwise;} \qquad i=1,2,...,m,
                \end{cases}
            \end{equation}
            and
            \begin{equation}\label{kinetic_ij-2}
                \begin{cases}
                    \begin{aligned}
                        \dot{c}_{ij}=\lambda_{ij}X_{ij} \quad \text{if}\;
                        &\{X_{ij}>0 \;\&\; c_i<1 \;\&\; c_j>0\}\qquad j \rightarrow i\\
                        \text{or}\; &\{X_{ij}<0 \;\&\; c_i>0 \;\&\; c_j<1\} \qquad i \rightarrow j
                    \end{aligned}\\
                    \dot{c}_{ij}=0 \qquad \qquad \qquad \qquad \qquad \quad \text{otherwise;} \qquad i,j = 1,2,...,m,
                \end{cases}
            \end{equation}

            Kinetic \Cref{kinetic_i0} in the more explicit form is $\dot{c}_{i0}=\lambda_{i0}(W_{i0}
                            -( \psi^{\theta}_M-\psi^{\theta}_A) -\mathcal{A}(1-2c) -k_{i-0}) $, where $W_{i0}
                            -( \psi^{\theta}_M-\psi^{\theta}_A)$ are the classical equilibrium thermodynamic terms and $ -\mathcal{A}(1-2c) -k_{i-0}$
            characterize deviation from the classical thermodynamic equilibrium, i.e., thermodynamic hysteresis.
            For Si I$\leftrightarrow$Si II phase transformation, complex hysteretic behavior depending on the principle stresses \citep{zarkevich2018lattice,Levitas2017LatticeCriterion,Levitas2017Triaxial-Stress-InducedPhases} implied that $k_{i-0}$ should depend on the principle stresses and volume fraction of variants ~\citep{babaei2020finite}.
            For Zr, such dependence is unknown, and the simplest	form of $k_{M-A}$, e.g., $k_{M-A}=const$ can be used.
            Calibration of $\mathcal{A}$ and $k_{M-A}$ can be made based on pressure for initiation of the direct and reverse transformations (i.e., at $c=0$ and $c=1$),  and if we assume that deviations of transformation pressure from the phase equilibrium pressure for direct and reverse transformations are the same, then we have only one condition for $-\mathcal{A} -k_{M-A} $.

            That is why for simulations we chose $k_{M-A}=0$ and will use  \Cref{kinetic_i0-2}.
            Since usually the threshold for the variant-variant transformation is smaller than for the austenite-martensite PT, we have to assume in simulations $k_{i-j}=0$ and use   \Cref{kinetic_ij-2}.
            For any of the above cases, the rate of change of the volume fraction
            $ c_i $  is determined by summation over all occurring transformations:
            \begin{equation}
                \dot{c}_i=\sum_{j=0}^{m}\dot{c}_{ij}; \qquad i=1,2,\dots,m,\quad j=0,1,\dots,m.
            \end{equation}

        \subsection{Mechanical equations}

            The thermodynamically consistent elasticity rule is
            \begin{eqnarray}\label{el-rule}
                \bs{\sigma} =J^{-1} \bs{P} \bs{\cdot}\bs{F}^{T}= J_e^{-1}  \bs{F}_e \bs{\cdot} \frac{\partial \psi^e}{\partial \bs{E}_e}\bs{\cdot}\bs{F}_e^{T}= J_e^{-1}  \bs{F}_e \bs{\cdot} ({\fg C}\bs{:} \bs{E}_e)\bs{\cdot}\bs{F}_e^{T} ;
                \qquad J_e=\frac{d V}{dV_t}=\det{\bs F_e}.
            \end{eqnarray}
            The mechanical equilibrium equations in the undeformed and deformed configurations are
            \begin{eqnarray}\label{eq-eq}
                \fg \nabla_0 \bs{\cdot} \bs{P}=\fg 0;
                \qquad
                \fg \nabla \bs{\cdot} \fsg=\fg 0.
            \end{eqnarray}
            For small strains, these equations simplify to the Hooke's law and traditional equilibrium equation
            \begin{eqnarray}\label{el-rule-small}
                \bs{\sigma} = \frac{\partial \psi^e}{\partial \fvep_e}= {\fg C}\bs{:} \fvep_e;
                \qquad \fg \nabla \bs{\cdot} \fsg=\fg 0.
            \end{eqnarray}

        \subsection{Macroscopic parameters}\label{Macroscopic}
            Macroscopic 
            first Piola-Kirchhoff stress, the deformation gradient, transformation strain, and volume fraction of  $\omega$ phase and each martensitic variant, averaged over the sample, are defined as (\cite{Hill1984OnStrain,Levitas1996SomeSurfaces,Petryk1998MacroscopicTransformation})
            \begin{equation}
                \bar{\fg P}=\frac{1}{V_0}\int_{V_0} {\fg P} dV_0;  \qquad
                \bar{\fg F}=\frac{1}{V_0}\int_{V_0} \fg F dV_0; \qquad
                \bar{\fg F}_t\simeq\frac{1}{V_0}\int_{V_0} \fg F_t dV_0;
                \qquad  \bar{\fvep}_t \simeq \frac{1}{V_0}\int_{V_0} {\fvep}_t dV_0;
                \label{av-Fp}
            \end{equation}
            \begin{equation}
                \bar{c}=\frac{1}{V_0}\int_{V_0} c dV_0;
                \qquad  \bar{c}_i=\frac{1}{V_0}\int_{V_0} c_i dV_0.
                \label{av-c}
            \end{equation}
            \begin{equation}
                \bar{\fsg}=(\det\bar{\fg F})^{-1}  \bar{\bs{P}} \bs{\cdot}\bar{\bs{F}}^T;
                \qquad
                \bar{\bs{E}}=\frac{1}{2}(\bar{\bs{F}}^T{\bs{\cdot}}\bar{\bs{F}}-\bs{I}).
                \label{av-sg-E}
            \end{equation}
            For  the first Piola-Kirchhoff stress, the deformation gradient, and volume fractions $c$ and $c_i$, averaging equations are strict; for the transformation deformation gradient and strain, they are approximate because, in the unloaded stress-free state, they generally are not compatible due to residual elastic strain.
            For small strains \Cref{av-Fp,av-sg-E} simplify to
            \begin{equation}
                \bar{\fsg }=\frac{1}{V}\int_{V} {\fsg } dV;  \qquad
                \bar{\fvep} = \frac{1}{V}\int_{V} {\fvep} dV;
                \qquad  \bar{\fvep}_t \simeq \frac{1}{V}\int_{V} {\fvep}_t dV.
                \label{av-small}
            \end{equation}

    \section{Material  parameters}\label{Model parameters} 
        The necessary material parameters for the model implementation, as presented in \Cref{model}, are provided in \Cref{lattice} and \Cref{tab1}.
        The $\alpha\rightarrow\omega$ transformation in Zr ( hcp$\rightarrow$simple hexagonal) follows the Silcock transformation pathway~\cite{wenk2013orientation}.
        In the coordinate system 1,2,3 (see \Cref{1pbc_x_c}), the transformation strains for  three $\omega$ variants are determined  in terms of the lattice parameters ($a$ and $c$) of $\alpha$ and $\omega$ phases and the rotation tensor $\fg \Omega$ by $60^o$ around axis 3 as reported in~\cite{yeddu2016alpha}:
        \begin{equation}
            {\fvep}_{t1}=
            \begin{pmatrix}
                \frac{c_\omega-a_\alpha}{a_\alpha} & 0 & 0 \\
                0 & \frac{2\sqrt{3}a_\omega-3\sqrt{3}a_\alpha}{3\sqrt{3}a_\alpha} & 0 \\
                0 & 0 & \frac{a_\omega-c_\alpha}{c_\alpha}
            \end{pmatrix}; \quad
            {\fvep}_{t2}=  \fg \Omega \cdot {\fvep}_{t1} \cdot \fg \Omega^T;  \quad
            {\fvep}_{t3}=  \fg \Omega \cdot {\fvep}_{t2} \cdot \fg \Omega^T; \quad
            \fg \Omega=\begin{pmatrix}
                \frac{1}{2} & \frac{\sqrt{3}}{2} &0 \\
                \frac{\sqrt{3}}{2} & \frac{1}{2}& 0 \\
                0	& 0 &1
            \end{pmatrix}. \label{epsten}
        \end{equation}
        Here, $c$ axes of $\alpha$ and $\omega$ phases are along the axes 3 and 1, respectively.
        The lattice parameters for $\alpha$ and $\omega$ phases of Zr, as measured by \citet{Pandey-Levitas-2020}, are given in \Cref{lattice}.
        Based on \Cref{epsten}, matrices of the components of tensors  ${\varepsilon}_{t2}$ and ${\varepsilon}_{t3}$ in the coordinate system 1,2,3  are related by ${\varepsilon}_{t2}= \Omega {\varepsilon}_{t1} \Omega^T$ and ${\varepsilon}_{t3}= \Omega {\varepsilon}_{t2} \Omega^T$.
        Then the transformation strain components  for all three variants are as follows:
        \begin{align}
            \scalebox{0.9}{$
                {\fvep}_{t1}=
                \begin{pmatrix}
                    -0.0318 & 0 & 0 \\
                    0 & 0.038 & 0 \\
                    0 & 0 & -0.022
                \end{pmatrix};
                \quad
                {\fvep}_{t2}=
                \begin{pmatrix}
                    0.0206 & 0.0302 & 0 \\
                    0.0302 & -0.0144 & 0 \\
                    0 & 0 & -0.022
                \end{pmatrix};
                \quad
                {\fvep}_{t3}=
                \begin{pmatrix}
                    0.0206 & -0.0302 & 0 \\
                    -0.0302 & -0.0144 & 0 \\
                    0 & 0 & -0.022
                \end{pmatrix},
            $}
            \label{epst}
        \end{align}
        with the transformational ratio of volumes $J_t=0.9847$ and the transformation volumetric strain ${\vep}_{tv}= J_t-1=-0.0153$.
        In the small strain approximation  ${\vep}_{t}^v=\tr {\fvep}_{t}=-0.0158$.
        To separate the effect of the pressure and deviatoric stress tensor, it is convenient to split ${\fvep}_{t}$ into spherical  ${\fvep}_{t0}= 1/3 {\vep}_{tv}  \fg I= -0.0053 \fg I$ and deviatoric $\fg e_{ti}$ parts
        \begin{align}
            \scalebox{0.8}{$
                {\fg e}_{t1}=
                \begin{pmatrix}
                    -0.0371 & 0 & 0 \\
                    0 & 0.0327 & 0 \\
                    0 & 0 & -0.0273
                \end{pmatrix};
                \quad
                {\fg e}_{t2}=
                \begin{pmatrix}
                    0.0153 & 0.0302 & 0 \\
                    0.0302 & -0.0197 & 0 \\
                    0 & 0 & -0.0273
                \end{pmatrix};
                \quad
                {\fg e}_{t3}=
                \begin{pmatrix}
                    0.0153 & -0.0302 & 0 \\
                    -0.0302 & -0.0197 & 0 \\
                    0 & 0 & -0.0273
                \end{pmatrix}.
            $}
            \label{epst-dev}
        \end{align}

        \begin{table}[h]
            \centering
            \caption{Lattice parameters for $\alpha$ and $\omega$ phases}
            \begin{tabular}{c c c c c}
                \hline
                Parameter  & $a_\alpha$ & $c_\alpha$  & $a_\omega$ & $c_\omega$ \\
                \hline
                Length (\AA) & 3.2317 & 5.1450 & 5.0320 & 3.1289 \\
                \hline
            \end{tabular}
            \label{lattice}
        \end{table}
        The elastic constants for both phases are obtained from \citet{wang2012first}.
        The constants $C_\alpha^{ij}$ in \Cref{tab1} represent the independent elastic constants of the austenite, while ${C}_\omega^{ij}$ denote those of the   martensitic variant 1, both in the local crystallographic axes.
        The components of the tensor of elastic moduli $C^{ijkl}$ in the axes 1,2,3 are calculated using the following expressions:
        \begin{eqnarray}
            \nonumber
            C^{ijkl}=\sum_{n=1}^{3}[\lambda^n\delta^{in}\delta^{jn}\delta^{kn}\delta^{ln}+\mu^n(\delta^{in}\delta^{jn}
            \delta^{kl}+\delta^{ij}\delta^{kn}\delta^{ln})\\
            +\nu^n(\delta^{in}\delta^{jk}\delta^{ln}+\delta^{jn}\delta^{ik}\delta^{ln}+\delta^{in}\delta^{jl}\delta^{kn}
            +\delta^{jn}\delta^{il}\delta^{kn})],
        \end{eqnarray}
        where $\lambda^n$, $\mu^n$ and $\nu^n$ for a hexagonal crystal lattice is given by \Cref{hexagonal}, respectively:
        \begin{align}{\label{hexagonal}}
            \nonumber
            &\lambda^1=\lambda^2=0, \\\nonumber
            &\lambda^3=C^{11}+C^{33}-2(C^{13}+C^{44}),\\\nonumber
            &2\mu^1=2\mu^2=C^{12}, \qquad \;\;2\mu^3=2C^{13}-C^{12},\\
            &4\nu^1=4\nu^2=C^{11}-C^{12},\qquad \;\; 4\nu^3=4C^{44}-C^{12}-C^{11}.
        \end{align}
        Matrices of the components  of the tensor of elastic moduli  for variants 2 and 3 in the coordinate system 1,2,3 are determined  by $ C_{2}={{\Omega (\Omega C_{1} \Omega}^T)\Omega^T}$ and $ C_{3}={\Omega (\Omega C_{2} \Omega}^T)\Omega^T$.

        Under hydrostatic conditions, the phase equilibrium pressure $p^{eq}_0=3.4$ GPa \citep{velisavljevic2011effects}.
        The jump in the thermal energy $\Delta\psi^\theta = -p^{eq}_0(J_t-1)=0.052$ GPa, where we took into account   that the contribution of the elastic strain and changes in elastic moduli can be neglected.
        To reproduce a discrete microstructure, strain localization is necessary.
        As noted in~\cite{babaei2020finite}, to achieve strain localization, the strain rate should be commensurate with the rate of transformation.
        The kinetic coefficient $\lambda$ and interaction parameter $\mathcal{A}$ are chosen to satisfy this condition.

        \begin{table}[h]
            \centering
            \caption{
                Material parameters including kinetic coefficient $\lambda \, \rm{(Pa \cdot s)}^{-1}$,
                interaction coefficient $\mathcal{A}$, jump in the thermal energy
                $ \Delta \psi^{\theta} $,  and elastic constants, all in GPa.  
            }
            \begin{tabular}{c c c c c c c c c }
                \hline
                $\lambda$   & $\mathcal{A}$ & $\Delta\psi^\theta $  				\\
                \hline
                5$\times10^{-3}$ & 0.028 & 0.052  \\
                \hline
                $ C_\alpha^{11}$  & $C_\alpha^{12}$ & $C_\alpha^{13}$ & $C_\alpha^{33}$ & $C_\alpha^{44}$ 	\\
                \hline
                146.7 & 68.5 & 71.0 & 163.3 & 26.0 \\
                \hline
                $ C_\omega^{11}$  & $C_\omega^{12}$ & $C_\omega^{13}$ & $C_\omega^{33}$ & $C_\omega^{44}$ 	\\
                \hline
                161.7 & 72.6 & 53.5 & 195.6 & 33.7 \\
                \hline
            \end{tabular}
            \label{tab1}
        \end{table}

    \section{Phase transformation start and finish criteria in the stress space}\label{PT-criteria}

        In small strain approximation, explicit equations for criteria for PT between $A$ and $M_i$  utilizing \Cref{Ws-sm} for the transformation work are
        \begin{align}\label{A-Mi}
            A \rightarrow M_i \qquad  X_{i0} \rvert_{c=0}= \fsg \bs{\cdot}  \bs{:}\bs{\varepsilon}_{ti}
            -\Delta \psi^{\theta} -\mathcal{A}\geq k_{M-A}   ;\\
            M_i \rightarrow A \qquad  X_{i0} \rvert_{c=1}= \fsg \bs{\cdot}  \bs{:}\bs{\varepsilon}_{ti}
            -\Delta \psi^{\theta} +\mathcal{A}\leq - k_{M-A}   .
        \label{Mi-A}
        \end{align}
        It is clear that $A$ and $k_{M-A}$ cannot be separated experimentally and can be combined.
        In explicit form, these criteria look as follows
        \begin{align}	\label{A-M1}
            A \rightarrow M_1 \qquad		-0.0318 \sg^{11}+ 0.038 \sg^{22}-0.022  \sg^{33} \geq 0.052 +\mathcal{A}+k_{M-A};\\
            M_1 \rightarrow A \qquad		-0.0318 \sg^{11}+ 0.038 \sg^{22}-0.022  \sg^{33} \leq 0.052 -\mathcal{A} - k_{M-A};
        \label{M1-A}
        \end{align}
        \begin{align}
            A \rightarrow M_2 \qquad	0.026  \sg^{11}-0.0144 \sg^{22}-0.022  \sg^{33} \geq 0.052 +\mathcal{A} + k_{M-A}-0.0604\sg^{12};\\
            M_2 \rightarrow A \qquad		0.026  \sg^{11}-0.0144 \sg^{22}-0.022  \sg^{33} \leq 0.052 -\mathcal{A}-k_{M-A} -0.0604 \sg^{12};
        \label{A-M2}
        \end{align}
        \begin{align}
            A \rightarrow M_3 \qquad	0.026  \sg^{11}-0.0144 \sg^{22}-0.022  \sg^{33} \geq 0.052 +\mathcal{A} +k_{M-A}+0.0604\sg^{12};\\
            M_3 \rightarrow A \qquad		0.026  \sg^{11}-0.0144 \sg^{22}-0.022  \sg^{33} \leq 0.052 -\mathcal{A}-k_{M-A} +0.0604 \sg^{12}.
            \label{M3-A}
        \end{align}
        Parameter $A+k_{M-A}$ depends on the preliminary plastic deformation and resultant microstructure ~\cite{Kumaretal-Acta-20,Kumaretal-Scripta-24,Pandey-Levitasetal-JAP-Zr-24}.
        Both $\Delta \psi^{\theta}$ and $A$ depend on impurities~\cite{velisavljevic2011effects}.
        The  terms with shear stresses is moved to the right side to simplify analysis of the PT criteria in the space of normal stresses.
        The hystereses of the PT in terms of the thermodynamic driving force is
        \begin{align}\label{hystereses}
            X_{i0} \rvert_{c=0}-X_{i0} \rvert_{c=1}= (\fsg\rvert_{c=0}-\fsg\rvert_{c=1}) \bs{\cdot}  \bs{:}\bs{\varepsilon}_{ti}=
                 2(	\mathcal{A}+k_{M-A})\geq 0,
        \end{align}
        which should be applied to all specific \Cref{A-M1}--\Cref{M3-A}.
        If some normal or shear stress is kept the same during the initiation of the direct  and reverse PT, they disappear the expression for the hysteresis and do not affect
        the hysteresis in other stresses.
        If , e.g., shear stress  $\sg^{12}$ changes sign, it significantly reduces hysteresis for normal stresses, making it zero	at $\sg^{12}= \mathcal{A}/0.0602=0.465\, GPa$ (data from  \Cref{tab1} is used).

        Decomposing stress tensor into a spherical and deviatoric $\fg S$ parts, we obtain
        \begin{align}\label{p-dev}
            \fsg=p \fg I+\fg S \quad \rightarrow \quad \fsg  \bs{:}\bs{\varepsilon}_{ti}= -p \vep_{t}^v+\fg S \bs{:}\fg e_{ti};  \qquad
             \fsg  \bs{:}(\bs{\varepsilon}_{ti}-\bs{\varepsilon}_{tj})= \fsg  \bs{:}(\fg e_{ti}-\fg e_{tj})= \fg S  \bs{:}(\fg e_{ti}-\fg e_{tj}),
        \end{align}
        i.e., equations for the variant-variant transformations are independent of pressure.

        Then  \Cref{A-Mi} and  \Cref{Mi-A}
        can be resolved in the expression for the PT pressures versus deviatoric stress
        \begin{align}\label{A-Mi-p}
            A \rightarrow M_i \qquad  p\geq   \frac{\Delta \psi^{\theta} +\mathcal{A}+k_{M-A}-\fg S \bs{:}\fg e_{ti}}{|\vep_{tv}|}  ;\\
            M_i \rightarrow A \qquad p\leq   \frac{\Delta \psi^{\theta} -\mathcal{A}-k_{M-A}-\fg S \bs{:}\fg e_{ti}}{|\vep_{tv}|}.
        \label{Mi-A-p}
        \end{align}
        In explicit form these equations look like
        \begin{align}	\label{A-M1-p}
            A \rightarrow M_1 \qquad		p\geq   {3.2911 + 63.2911(\mathcal{A} + k_{M-A}) + 2.3481 S^{11} - 2.0696 S^{22} + 1.7278 S^{33}}  ;\\
            M_1 \rightarrow A \qquad		p\leq   {3.2911 - 63.2911(\mathcal{A} + k_{M-A}) + 2.3481 S^{11} - 2.0696 S^{22} + 1.7278 S^{33}};
        \label{M1-A-p}
        \end{align}
        \begin{align}
            A \rightarrow M_2 \qquad		p\geq   {3.2911 + 63.2911(\mathcal{A} + k_{M-A}) - 0.9683 S^{11} + 1.268 S^{22} + 1.7278 S^{33} - 3.8228 S^{12}}  ;\\
            M_2 \rightarrow A \qquad		p\leq   {3.2911 - 63.2911(\mathcal{A} + k_{M-A}) - 0.9683 S^{11} + 1.268 S^{22} + 1.7278 S^{33} - 3.8228 S^{12}}  ;
        \label{A-M2-p}
        \end{align}
        \begin{align}
            A \rightarrow M_3 \qquad		p\geq   {3.2911 + 63.2911(\mathcal{A} + k_{M-A}) - 0.9683 S^{11} + 1.268 S^{22} + 1.7278 S^{33} + 3.8228 S^{12}}  ;\\
            M_3 \rightarrow A \qquad		p\leq   {3.2911 - 63.2911(\mathcal{A} + k_{M-A}) - 0.9683 S^{11} + 1.268 S^{22} + 1.7278 S^{33} + 3.8228 S^{12}}.
            \label{M3-A-p}
        \end{align}

        Since $S^{11}+S^{22}+S^{33}=0$, one of the stress, e.g., $S^{33}$ can be excluded, leading to the following equations
        \begin{align}	\label{A-M1-p}
            A \rightarrow M_1 \qquad		p\geq   {3.2911 + 63.2911(\mathcal{A} + k_{M-A}) + 39.2532 S^{11} - 3.7975 S^{22}}  ;\\
            M_1 \rightarrow A \qquad		p\leq   {3.2911 - 63.2911(\mathcal{A} + k_{M-A}) + 39.2532 S^{11} - 3.7975 S^{22}};
        \label{M1-A-p}
        \end{align}
        \begin{align}
            A \rightarrow M_2 \qquad		p\geq   {3.2911 + 63.2911(\mathcal{A} + k_{M-A}) - 2.6962 S^{11} - 0.4810 S^{22} - 3.8228 S^{12}}  ;\\
            M_2 \rightarrow A \qquad		p\leq   {3.2911 - 63.2911(\mathcal{A} + k_{M-A}) - 2.6962 S^{11} - 0.4810 S^{22} - 3.8228 S^{12}}  ;
        \label{A-M2-p}
        \end{align}
        \begin{align}
            A \rightarrow M_3 \qquad		p\geq   {3.2911 + 63.2911(\mathcal{A} + k_{M-A}) - 2.6962 S^{11} - 0.4810 S^{22} + 3.8228 S^{12}}  ;\\
            M_3 \rightarrow A \qquad		p\leq   {3.2911 - 63.2911(\mathcal{A} + k_{M-A}) - 2.6962 S^{11} - 0.4810 S^{22} + 3.8228 S^{12}}.
            \label{M3-A-p}
        \end{align}

        {The critical shear stresses for $\alpha$ Zr for prismatic and pyramidal glide range from 35 to 40 MPa and from 55 to 60 MPa, respectively; for hcp pyramidal slip they range from 90 to 110 MPa     ~\cite{poty2011classification}.
        Taking for an estimate $S^{12}=0.10 $ GPa and $S^{22}=0.17 $ GPa, we obtain from Eqs. (\ref{M1-A-p}) and (\ref{M3-A-p})  the reduction in PT pressure by   0.38 and  0.65 GPa, respectively.
        For strongly predeformed single crystals, these number can be essentially higher.
        For a severely deformed polycrystal at 2 GPa, the yield strength in compression is 1.20 GPa and shear 0.69 GPa~\cite{Levitas-etal-NatCom-23}.
        Using a Taylor factor for hcp crystal of 3~\cite{Caceres-PhilMag-08}, we estimate $S^{12}=0.23 $ and $S^{22}=0.40 $ GPa.
        Then for the grain of such a polycrystal, the reduction in the PT pressure is by  0.88 and  1.52 GPa, respectively.
        While these numbers are quite high, they are not sufficient to describe the reduction in PT pressure
        from 6.0 to 0.67 GPa observed during plastic compression in~\cite{Linetal-MRL-23,Lin-Levitas-ResSquare-22} and even from 5.4 to 2.7 GPa in~\cite{Levitas-etal-NatCom-23}. 
        This leads to the conclusion that the reduction is mostly due to the plastic strain-induced PT mechanisms,
        i.e., nucleation at the tip of defects generated during plastic flow, like dislocation pileups or twins~\cite{Levitas-PRB-2004,levitas2018high,Levitas_2019,Levitas-IJP-21,Levitas-MT-23}.

        For variant-variant transformations,  the thermodynamic equilibrium equations looks like
        \begin{align}\label{Mi--Mj}
            M_j \rightarrow M_i	\qquad		W_{ij}= \fsg  \bs{:}(\bs{\varepsilon}_{ti}-\bs{\varepsilon}_{tj}) \geq k_{j-i};\\
            M_i \rightarrow M_j	\qquad		W_{ij}= \fsg  \bs{:}(\bs{\varepsilon}_{ti}-\bs{\varepsilon}_{tj}) \leq - k_{i-j}.
        \end{align}
        Due to equivalence of martensitic variants, we have $k_{j-i}=k_{i-j}=k_{M-M}$ for all $i$ and $j$.
        In explicit form, these conditions can be presented as
        \begin{align}	\label{M1-M2-p}
            &M_1 \rightarrow M_2 \qquad      0.0524 S^{11} -0.0524 S^{22} +0.0604 S^{12} \geq k_{M-M};\\
            &M_2 \rightarrow M_1 \qquad		   0.0524 S^{11} -0.0524 S^{22} +0.0604 S^{12}\leq - k_{M-M};
        \end{align}
        \begin{align}
            &M_2 \rightarrow M_3 \qquad     0.1208 S^{12} \leq  -k_{M-M} ;\\
            &M_3 \rightarrow M_2 \qquad		0.1208 S^{12}\geq k_{M-M} ;
        \label{M2-M3-p}
        \end{align}
        \begin{align}
           &M_1 \rightarrow M_3 \qquad		  {0.0524 S^{11} -0.0524 S^{22} - 0.0604 S^{12}} \geq k_{M-M}; \\
            &M_3 \rightarrow M_1 \qquad		   {0.0524 S^{11} -0.0524 S^{22} -0.0604 S^{12}}  \leq - k_{M-M} .
        \label{M3-M1-p}
        \end{align}}

        According to \Cref{p-dev}, all stresses in the above equations can be substituted with the deviatoric stresses.

        In geometric interpretation in 3D space of the components of normal stresses, each of the conditions  \Cref{A-M1}- \Cref{M3-A} is represented
        by planes orthogonal to the corresponding $\fvep_{ti}$ vector (without shear components).

    \section{Boundary Conditions and  Analytical  Solutions for Equilibrium Phase Transformation}\label{analytical}

        Six different boundary conditions are used to study the evolution of microstructure of the $\omega$ phase.
        \begin{enumerate}
            \item Set 1 -- periodic boundary conditions in 1 direction and compressive strain loading in direction 1; zero tractions at faces orthogonal to the axes 2 and 3.
            \item Set 2 -- periodic boundary conditions in all directions with compressive strain loading in direction 1.
            \item Set 3 -- symmetric boundary conditions on three perpendicular faces with compressive strain loading in direction 1 and zero tractions at faces orthogonal to the axes 2 and 3.
            \item Set 4 -- periodic boundary conditions in all directions with compressive strain loading in direction 3.
            Sets 2 and 4 mimic the boundary conditions for shock loadings in direction 1 and 3, respectively.
            \item Set 5 -- hydrostatic stress loading (like within gas or liquid).
            \item Set 6 -- complex loading of selected two grains in a polycrystal with 30 grains; the polycrystal sample is subjected to periodic boundary conditions and compressive strain loading in direction 1.
        \end{enumerate}

        For each of these boundary conditions but in set 6   we first  analyze analytical homogeneous solution.
        The analytical plots are obtained  in three parts:
        \begin{enumerate}[(i)]
            \item \textbf{Elastic Austenite:} During any loading, the starting austenite deforms elastically leading to a straight line till PT initiation.
            The stress and strain corresponding to PT initiation are evaluated from the condition $X_{i0}=0$ at $c=0$, where  $X_{i0}$ is given in \Cref{Xs} for small strains and \Cref{dissipation} for finite strains, and the elasticity rule \Cref{el-rule} with $\bs{C}=\bs{C_0}$.
            The total strain $\bs E=\bs E_e$ as there is no PT contribution here.
            For small strain approximation, the elasticity rule is  $\bs{\sigma} = {\fg C}_0\bs{:} \fvep_e$ and $\fvep=\fvep_e$.
            \item \textbf{Phase Transformation:} During the PT, the stress and strain are represented in a parametric form as a function of $c$ (volume fraction of martensite) and plotted for $c\in [0,1]$.
            Thus, the stress and elastic strain are calculated from the condition $X_{i0}=0$ for arbitrary $c\in [0,1]$, where  $X_{i0}$ is defined in \Cref{Xs} for small strains and \Cref{dissipation} for finite strains, and elasticity rule \Cref{el-rule}
                with $\bs{C}$ from  \Cref{elastic_energy}.
                The total strain is the calculated as $\bs E = \bs F_t\bs\cdot\bs E_e \bs\cdot \bs F_t + \bs E_t$.
                For small strains,  the elasticity rule is  $\bs{\sigma} = {\fg C}\bs{:} \fvep_e$ and $\fvep=\fvep_e+\fvep_t$.
            \item \textbf{Elastic Martensite:} After the austenite completely transforms to martensite, elastic deformation of martensite is observed.
            The stress and elastic strain are calculated like in part (i) but with properties of martensite.	The total strain is the calculated as $\bs E = \bs F_t\bs\cdot\bs E_e \bs\cdot \bs F_t + \bs E_t$ for  $c=1$.
            For small strains,  the elasticity rule is  $\bs{\sigma} = {\fg C}\bs{:} \fvep_e$ and $\fvep=\fvep_e+\fvep_t$ for  $c=1$.
        \end{enumerate}
        Below, we will derive analytical expressions for Sets 1 to 5.
        Corresponding plots and comparison with numerical solutions will be presented in \Cref{results}.

        \subsection{Sets 1 and  3:}\label{1and3}
            For zero lateral stresses, it is evident that  the first variant maximizes the transformation work $W_{i0}$ and  the thermodynamic driving force  $X_{i0}$.
            That is why the first variant will be considered only, which is confirmed below by FEM simulations.
            For prescribed stresses, the corresponding $\bs\varepsilon^e$ is calculated as
            \begin{equation}
                \bs\varepsilon_e = \bs{G}:\bs\sigma,
                \label{Hookes}
            \end{equation}
            where \bs{G} is the effective compliance of austenite-martensite mixture defined using mixture theory as
            \begin{equation}
             {\bs G}^{-1}={1-c}{\bs{G_A}^{-1}}+{c}{\bs{G_M}^{-1}}; \qquad \bs{C}=(1-c)\bs{C_A}+c\bs{C_M}.
                \label{equiG}
            \end{equation}
            { Using \Cref{equiG} to calculate the analytical solution leads to a nonlinear equation dependent on higher powers of $c$ and nonlinear stress-strain curves; that is why it will be called below a "nonlinear" model, while it is basically just $c$-dependent moduli.
            These equations become cumbersome and difficult to solve without numerical approximations.
            Hence, Mathematica is used to solve these equations and plot them in the figures shown in the following section.}

            For simplicity and to present the analytical procedure followed, we neglect $c$-dependence of \bs{G} and use elastic properties of austenite.
            In all of our analysis, we use Voigt notation where $\bs G$ is represented as a 6$\times$6 matrix, $\bs \sigma$ and $\bs \varepsilon$ are represented as 6$\times$1 vector.
            Hence, $\sigma_{ii}$ and $\varepsilon_{ii}$ are represented as $\sigma_i$ and $\varepsilon_i$ for compactness.
            Since only $\sigma_{1}\neq0$, we can simplify \Cref{Hookes} to
            \begin{equation}
                \varepsilon^{i}_e = {G}_{i1}\sigma_{1}, \qquad i=1,2,3.
                \label{Hookes2}
            \end{equation}
            The total strain $\bs\varepsilon$ becomes
            \begin{equation}
                \varepsilon^{i} = \varepsilon^{i}_e+c\varepsilon^{i}_{t1}.
                \label{ts}
            \end{equation}
            The phase equilibrium condition for the driving force is
            \begin{align}
                X_{10} =
                \bs\sigma\bs{:}\bs\varepsilon_t-\Delta\psi-A(1-2c) = 0,
                \label{df_limit-0}
            \end{align}
            where $\bs\sigma\bs{:}\bs\varepsilon_t=\varepsilon^{1}_{t1}\sigma^{1}$.
            Eliminating $c$ from \Cref{df_limit-0} using \Cref{Hookes2,ts}, we obtain a thermodynamically equilibrium relation between $\sigma^{1}$ and $\varepsilon^{i}$:
            \begin{align}
                \sigma^{1}=\frac{-2A}{\varepsilon^{1}_{t1}\varepsilon^{i}_{t1}-2AG^{i1}}\varepsilon^{i}+\frac{(\Delta\psi+A)}{\varepsilon^{1}_{t1}\varepsilon^{i}_{t1}-2AG^{i1}}\varepsilon^{i}_{t1}.
                \label{df2}
            \end{align}
            In the loading direction 1,	\Cref{df2} describes a decreasing stress-strain function, i.e., a strain softening, if the material parameter $A$ satisfies the inequality
            \begin{equation}
                A<(\varepsilon^{1}_{t1})^2/2G^{11},
            \end{equation}
            {which is the case for Zr.} We took into account that $A>0$.
            Since for all the boundary conditions there are no shear stresses, we do not consider them in our analysis.

        \subsection{Set 2:}
            {It follows from the boundary conditions in Set 2 that
            \begin{align}
                &\varepsilon^{1}=	 \varepsilon^{1}_e+ c \varepsilon^{1}_{t1};
                \label{str-set2-1}\\
                &\varepsilon^{i}=	 \varepsilon^{i}_e+ c \varepsilon^{i}_{t1}=0,  \;\; i=2,3 \quad \implies \quad  \varepsilon^{2}_e=-c \varepsilon^{2}_{t1}; \quad
                \varepsilon^{3}_e=- c \varepsilon^{3}_{t1}.
                \label{str-set2}
            \end{align}
            We took into account  that  the first variant, maximizing the transformation work $W_{i0}$ and  the thermodynamic driving force  $X_{i0}$, is only present, which is confirmed below by FEM simulations.
             The corresponding $\bs\sigma$ can be calculated as}
            \begin{equation}
                \bs\sigma = \bs{C}(c):\bs\varepsilon_e; \qquad 	\sigma^{i} = {C}^{ij}(c)\varepsilon^{i}_e\qquad i=1,2,3,
                \label{Hookes_2}
            \end{equation}
            As noted in \Cref{1and3}, for simplicity, in analytical solution presented here we neglect $c$-dependence of \bs{C} and use elastic properties of austenite.
            The analytical plots presented are generated using Mathematica using the mixture theory based effective elasticity tensor.

            The phase equilibrium condition for the driving force is
            \begin{align}
                X_{10} = \bs\sigma\bs{:}\bs\varepsilon_{t1}-\Delta\psi-A(1-2c) =
                \varepsilon^{i}_{t1}\sigma^{i}-\Delta\psi-A(1-2c) = 0.
                \label{df_limit}
            \end{align}
            Eliminating $c$ from \Cref{str-set2-1,Hookes_2,df_limit}, we obtain
            \begin{equation}
                \sigma^1 = \frac{K_1\varepsilon^1+L_1}{M},
                \label{s1}
            \end{equation}
            where
            \begin{align}
                K_1 = &C^{11}(-2 A  +  C^{22}  (\varepsilon^{2}_{t1})^2 + 2C^{23}\varepsilon^{2}_{t1}\varepsilon^{3}_{t1} +
                C^{33} (\varepsilon^{3}_{t1})^2 + \psi) - (C^{12}\varepsilon^{2}_{t1} + C^{13}\varepsilon^{3}_{t1})^2; \nonumber \\
                L_1 = &(A+\psi)(C^{11}\varepsilon^{1}_{t1} + C^{12} \varepsilon^{2}_{t1} + C^{13}\varepsilon^{3}_{t1});\nonumber \\
                M = &-2A+C^{11} (\varepsilon^{1}_{t1})^2+2 C^{12} \varepsilon^{1}_{t1} \varepsilon^{2}_{t1}+2 C^{13} \varepsilon^{1}_{t1} \varepsilon^{3}_{t1}+C^{22}
                   (\varepsilon^{2}_{t1})^2+2 C^{23} \varepsilon^{2}_{t1} \varepsilon^{3}_{t1}+C^{33} (\varepsilon^{3}_{t1})^2.
                   \label{s1_coeff}
            \end{align}
            Similarly, $\sigma^2$ and $\sigma^3$ can be expressed as a function of $\varepsilon^1$ as
            \begin{align}
                \sigma^2 = \frac{K_2\varepsilon^1+L_2}{M} ; \qquad
                \sigma^3 = \frac{K_3\varepsilon^2+L_3}{M},
                \label{s23}
            \end{align}
            where
            \begin{align}
                K_2 = &C^{12}(-2 A + C^{33} (\varepsilon^3_{t1})^2 + C^{12}\varepsilon^1_{t1} \varepsilon^2_{t1} + C^{13} \varepsilon^1_{t1} \varepsilon^3_{t1} + C^{23} \varepsilon^2_{t1} \varepsilon^3_{t1}) - (C^{11}\varepsilon^1_{t1} + C^{13}\varepsilon^3_{t1})(C^{22}\varepsilon^2_{t1} + C^{23} \varepsilon^3_{t1}); \nonumber\\
                L_2 = &(A+\psi) (C^{12} \varepsilon^1_{t1}+ C^{22} \varepsilon^2_{t1}+C^{23} \varepsilon^3_{t1}); \nonumber\\
                K_3 = &C^{13}(-2 A + C^{12}\varepsilon^1_{t1}\varepsilon^2_{t1} + C^{13}\varepsilon^1_{t1}\varepsilon^3_{t1} + C^{23}\varepsilon^2_{t1}\varepsilon^3_{t1} + C^{22}(\varepsilon^2_{t1})^2) - (C^{11}\varepsilon^1_{t1} + C^{12}\varepsilon^2_{t1}) (C^{23}\varepsilon^2_{t1} + C^{33}\varepsilon^3_{t1}); \nonumber \\
                   L_3 = &(A + \psi) (C^{13}\varepsilon^1_{t1} + C^{23}\varepsilon^2_{t1} + C^{33}\varepsilon^3_{t1}).
                   \label{s23_coeff}
            \end{align}

            For any of \Cref{s1,s23} to exhibit instability with a strain-softening, they should satisfy the inequality
            \begin{equation}
                \frac{K_i}{M}<0\qquad i=1,2,3
                \label{ineq}
            \end{equation}
      {For  Zr, $M>0$ and  $K_i>0$; hence, the \Cref{ineq} is not satisfied.}

        \subsection{Set 4:}
            Set 4 is very similar to set 2, except for a change in direction for the load application.
            Hence, instead of $\varepsilon_{11}\neq0$ we have $\varepsilon_{33}\neq0$.
            Because all the three variants appear as a result of having the same transformation strain in direction 3, we consider $c_1=c_2=c_3=c/3$ and
            \begin{eqnarray}
                \scalebox{0.9}{$
                    \fvep_t = \frac{1}{3}
                \sum_{i=1}^{3}\fvep_{ti}=
                            \begin{pmatrix}
                                0.0031 & 0 & 0 \\
                                0 & 0.0031 & 0 \\
                                0 & 0 & -0.022
                            \end{pmatrix};
                        $}
            \label{epst-4}
            \end{eqnarray}
            { Much smaller transformation strains in the lateral directions 1 and 2 explain much lower lateral resistance to the PT for periodic boundary conditions than for Set 2.} Next,
            \begin{align}
                &\varepsilon^{3}=	 \varepsilon^{3}_e+ \frac{c}{3} \varepsilon^{3}_{t};
                \label{str-set2-3}\\
                &\varepsilon^{i}=	 \varepsilon^{i}_e+ \frac{c}{3} \varepsilon^{i}_{t}=0,  \;\; i=1,2 \quad \implies \quad  \varepsilon^{1}_e=-\frac{c}{3} \varepsilon^{1}_{t}; \quad
                \varepsilon^{2}_e=- \frac{c}{3} \varepsilon^{2}_{t}.
                \label{str-set4}
            \end{align}
            The phase equilibrium condition for the driving force is
            \begin{align}
                X_{i0} = \bs\sigma\bs{:}\bs\varepsilon_{ti}-\Delta\psi-A(1-2c) =
                \varepsilon^{i}_{ti}\sigma^{i}-\Delta\psi-A(1-2c) = 0\qquad i=1,2,3.
                \label{df_limit_s4}
            \end{align}
            Since all the three variants appear at the same time, \Cref{df_limit_s4} can be averaged for the three variants and consider the total martensite as a singular variant with averaged properties (transformation strain and elastic modulus) of all the three variants.
            Using this simplification, \Cref{df_limit_s4} becomes
            \begin{align}
                X_{M\leftarrow A} = \varepsilon^{i}_{t}\sigma^{i}-\Delta\psi-A(1-2c) = 0\qquad i=1,2,3.
                \label{df_limit_s4_av}
            \end{align}
            Eliminating $c$ from \Cref{str-set4,Hookes_2,df_limit_s4_av}, we obtain
            \begin{equation}
                \sigma^i = \frac{P_i\varepsilon^i+Q_i}{M},
                \label{s3_4}
            \end{equation}
            where
            \begin{align}
                P_1 = &C^{13}(-2 A + C^{12}\varepsilon^1_t\varepsilon^2_t + C^{13}\varepsilon^1_t\varepsilon^3_t + C^{22}(\varepsilon^2_t)^2 + C^{23}\varepsilon^2_t\varepsilon^3_t) - (C^{11}\varepsilon^1_t + C^{12}\varepsilon^2_t)(C^{23}\varepsilon^2_t + C^{33}\varepsilon^3_t)\nonumber \\
                Q_1 = &(A+\psi)(C^{11} \varepsilon^1_t+C^{12} \varepsilon^2_t+C^{13} \varepsilon^3_t);\nonumber\\
                P_2 = &C^{23}(-2 A + C^{12}\varepsilon^1_t\varepsilon^2_t + C^{13}\varepsilon^1_t\varepsilon^3_t + C^{11}(\varepsilon^1_t)^2 + C^{23}\varepsilon^2_t\varepsilon^3_t) - (C^{12}\varepsilon^1_t + C^{22}\varepsilon^2_t)(C^{13}\varepsilon^1_t + C^{33}\varepsilon^3_t);\nonumber \\
                Q_2 = &(A+\psi)(C^{22} \varepsilon^2_t+C^{12} \varepsilon^1_t+C^{23} \varepsilon^3_t);\nonumber\\
                P_3 = &C^{33} (- 2 A + C^{11}   (\varepsilon^{1}_t)^2+ 2 C^{12}   \varepsilon^{1}_t \varepsilon^{2}_t+C^{22}  (\varepsilon^{2}_t)^2) - (C^{13}\varepsilon^{1}_t+C^{23}\varepsilon^{2}_t)^2 ;\nonumber\\
                Q_3 = &(A+\psi)(C^{33} \varepsilon^{3}_t  + C^{13} \varepsilon^{1}_t + C^{23}\varepsilon^{2}_t).
            \end{align}
            Parameter			$M$ in \Cref{s3_4} is defined in \Cref{s1_coeff}. \Cref{s3_4} will exhibit strain-softening only when the following inequality is satisfied
            \begin{equation}
                \frac{P_i}{M}<0\qquad i=1,2,3
                \label{ineq4}
            \end{equation}
            For set 4, all the three conditions \Cref{ineq4} are satisfied.

        \subsection{Set 5:}
            {For hydrostatic loading $\bs \sigma = -p \bs{I}$ in Set 5, it is sufficient to operate by volumetric strain and use \Cref{e26a}.
            Using \Cref{Hookes}, we get
            \begin{equation}
                \varepsilon_{i}^e = -(G_{i1}+G_{i2}+ G_{i3})p \quad \implies \quad
                \varepsilon_{v}^e=\Sigma_{i=1}^3 \varepsilon_{i}^e=k_{ {R}} p; \quad
                k_{ {R}}=  -(G_{11}+2G_{12}+ 2G_{13}+ G_{22}+2G_{23}+ G_{33}),
                \label{Hookes_5}
            \end{equation}
            where $k_{{R}}$	is the bulk compliance of a polycrystal in the Reuss approximation.	The phase equilibrium condition for the driving force  becomes
            \begin{equation}
                X_{10} = -A(1-2c) - \varepsilon^{v}_tp - \psi = 0.
                \label{df_limit_5}
            \end{equation}
            Eliminating $c$ from \Cref{df_limit_5,Hookes_5,e26a}, we get}
            \begin{equation}
                p = \frac{2A}{-2Ak_R+(\varepsilon^{v}_t)^2}\varepsilon^v - \frac{A+\psi}{-2Ak_R+(\varepsilon^{v}_t)^2}\varepsilon^{v}_t.
            \end{equation}
            The condition for the negative for the hydrostatic loading is
            \begin{equation}
                A<(\varepsilon^v_t)^2/2k_R,
                \label{ineq5}
            \end{equation}
       {which  is not met  for Zr.} However, both stress and strain reduce, i.e., there is material instability. 

    \section{Study of multivariant microstructure evolution}\label{results}
        The scale-free model is implemented using finite elements in the open-source FEM code deal.II~\citep{Bangerth2007Deal.IIALibrary}.
        8-noded 3D cubic linear elements with first-order interpolation, 8 integration points, and full integration are employed for the implementation.
        For such elements, the calculation of the volume-averaged value of any parameter $a$ in the undeformed configuration is obtained by summing the values of $a$ at each quadrature point and dividing it by the number of quadrature points.
        {The $\alpha$-Zr single crystal is oriented with $c$ axis ($[0001]$) along direction 3, $[11\bar{2}0]$ along direction 1, and 
         $[1\bar{1}00]$ along direction 2.
        }

        To investigate the influence of the number of elements on the microstructure and seek a mesh-independent solution for each boundary condition, the problem is solved using a range of element counts, starting from 512 (8$\times$8$\times$8) and increasing up to 262,144 (64$\times$64$\times$64) elements.
        Considering the position vector $\fg r$ and volume fractions $c_i$ as 6 primary independent variables, the total number of degrees of freedom is 48 times the number of elements, i.e., 24,576 and 12,582,912, respectively.

        \subsection{Set 1}\label{Set1}
            Periodic boundary conditions are utilized in only direction 1 and then subjected to a compressive strain along direction 1.
            The lateral directions 2 and 3 are stress free.
            To examine the quasi-static behavior of the material accurately, the loading is applied at a strain rate of $5\times 10^{-6} s^{-1}$ in the elastic regime and then reduced to $2\times 10^{-9} s^{-1}$ just before the transformation initiation.

\begin{figure}[htp]
                \centering
                \subfloat{\resizebox{150mm}{!}{\includegraphics[width=1.0\textwidth]{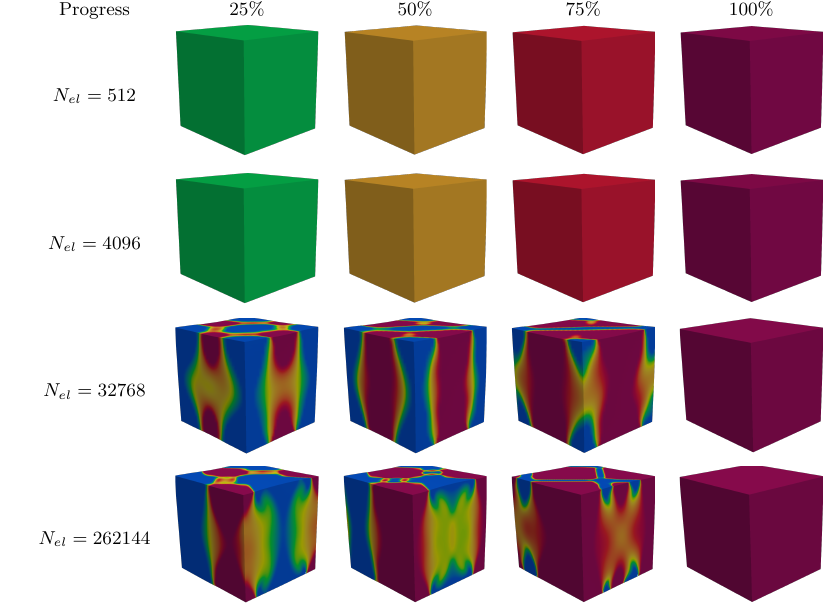}}}\hfill
                \subfloat{\hspace{10mm}\resizebox{20mm}{!}{\includegraphics[width=0.2\textwidth]{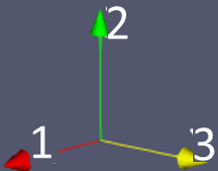}}}\hspace{35mm}
                \subfloat{\resizebox{60mm}{!}{\includegraphics[width=0.6\textwidth]{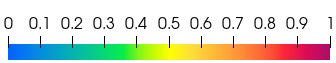}}}\hspace{30mm}
                \caption{Evolution of $\omega$ phase in a single crystal Zr under periodic boundary conditions in direction 1 and compressive strain in direction 1 with stress-free faces orthogonal to axes 2 and 3 (set 1) for varying number of elements.
                The $\omega$ phase  consists of variant 1 only. 
                The numbers at the top indicate different stages of completion of the simulation for different number of elements.}
                \label{1pbc_x_c}
            \end{figure}

            From the transformation strains given in \Cref{epst}, it is evident that  the first variant maximizes the transformation work $W_{i0}$ and  the thermodynamic driving force  $X_{i0}$.
            Indeed, under  compression  in direction 1,  only the first variant appears in the simulations. \Cref{1pbc_x_c} shows the evolution of volume fraction of the first variant at different stages of completion of the simulation for different number of elements.
            It can be seen that, the phase transformation progresses  homogeneously for  a smaller number of elements ($\leq 4,096$) only.
            For larger number of elements ($\geq 32,768$), although initially a homogeneous evolution is observed, non-homogeneous microstructure occurs clearly distinguishing austenite and martensite.
            This happens because phase interfaces, i.e., transition zone where $c$ varies between 0 and 1, are localized within 1-2 finite elements.
            They have higher energy due to the interaction energy  between austenite and martensite $\psi^{in} $ and energy of internal stresses due to large gradient in transformation strain in these regions.
            For small number of elements, volume of phase interfaces is relatively large and their appearance is energetically prohibited.

\begin{figure}[htbp]
                \centering
                \resizebox{70mm}{!}{\includegraphics{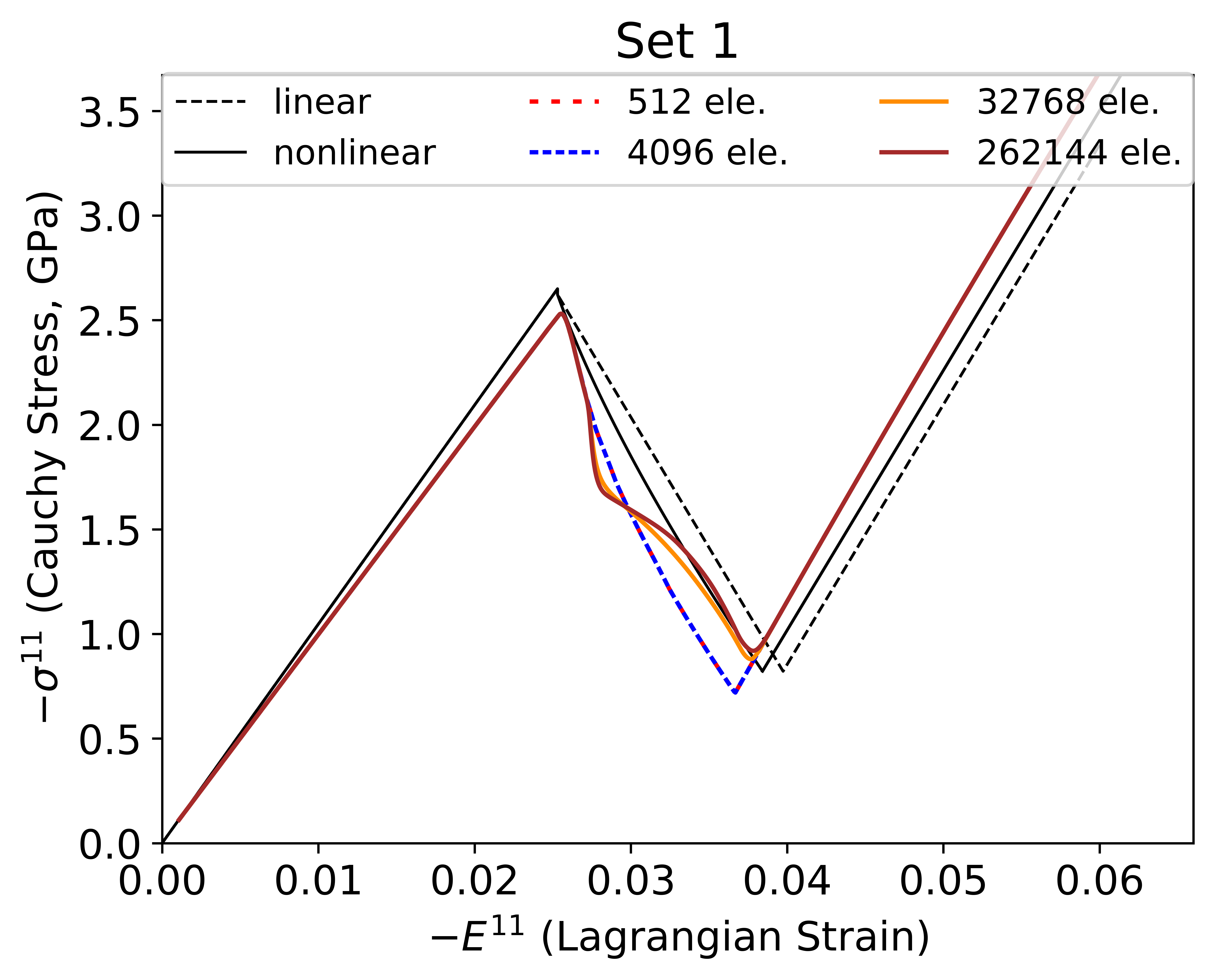}}
                \caption{Comparison between the analytical and finite element calculations with different number of elements for averaged Cauchy stress - Lagrangian strain plot for single crystal Zr under periodic boundary conditions and compressive strain in direction 1 with stress-free faces orthogonal to axes 2 and 3 (set 1).}
                \label{1pbc_x_p}
            \end{figure}

            \Cref{1pbc_x_p} shows the stress-strain ($\sigma^{11}$ vs. $E^{11}$) plots for different number of elements along with the analytical plot.
            {Analytically, the PT begins at a stress of $-2.650$ GPa and a strain of $-0.025$.
            The PT ends at a stress of $-0.822$ GPa and a strain of $-0.040$ for linear approach and a strain of $-0.038$ for the nonlinear calculation.
            In the simulations, the starting stress and strain exactly match for all the number of elements at $-2.529$ GPa and $-0.025$.
            The ending stress for smaller number of elements ($\leq 4,096$) is $-0.724$ GPa and the strain is $-0.037$, but for larger number of elements, the ending stress is $-0.954$ GPa and the strain is $-0.038$.
            The differences between the analytical and simulation results for smaller number of elements are due to the fact that the analytical solution uses a small strain formulation, as discussed in \Cref{analytical}, whereas the simulation uses a finite strain formulation.
            Also, as expected, the non-linear approach gives a closer match with the simulation results.
            Thus, the shape of the stress-strain curves with the smaller number of elements is similar to the analytical plots because of  uniform phase transformation.
            For larger number of elements, the kink in the stress-strain plots indicates the presence of heterogeneity.}

        \subsection{Set 2}\label{Set2}
            Periodic boundary conditions are utilized in all three directions and sample is subjected to a compressive strain along direction 1; averaged strains in directions 2 and 3  are zero,
            like in shock experiments.
            The loading is applied at a strain rate of $10^{-5} s^{-1}$ in the elastic regime and then reduced to $10^{-9} s^{-1}$ just before the material initiates transformation.

\begin{figure}[htp]
                \centering
                \subfloat{\resizebox{150mm}{!}{\includegraphics[width=1.0\textwidth]{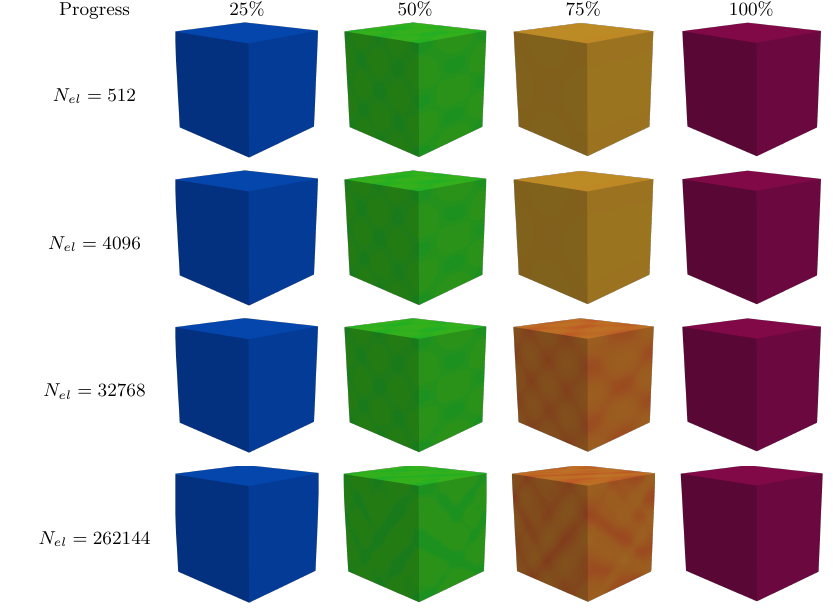}}}\hfill
                \subfloat{\hspace{10mm}\resizebox{20mm}{!}{\includegraphics[width=0.2\textwidth]{Images/axes.png}}}\hspace{35mm}
                \subfloat{\resizebox{60mm}{!}{\includegraphics[width=0.6\textwidth]{Images/colorbar_c.png}}}\hspace{30mm}
                \caption{Evolution of $\omega$ phase in single crystal Zr under periodic boundary conditions in all three directions and compressive strain in direction 1 for varying number of elements.
                The $\omega$ phase  consists of variant 1 only.
                The numbers at the top indicate different stages of completion of the simulation for different number of elements.}
                \label{3pbc_x_c}
            \end{figure}

            Like in  \Cref{Set1}, it is expected that only the first variant  appears when compressed in direction 1 and the same is observed in the simulations. \Cref{3pbc_x_c} shows the evolution of volume fraction of the first variant at different stages of completion of the phase transformation for varying number of elements.
            The phase transformation initially progresses homogeneously and at a later stage, a slightly heterogeneous microstructure starts to develop for $N_{el}\geq 32,768$.

\begin{figure}
                \centering
                \subfloat[\label{3pbc_x_px}]{\resizebox{70mm}{!}{\includegraphics[width=0.5\textwidth]{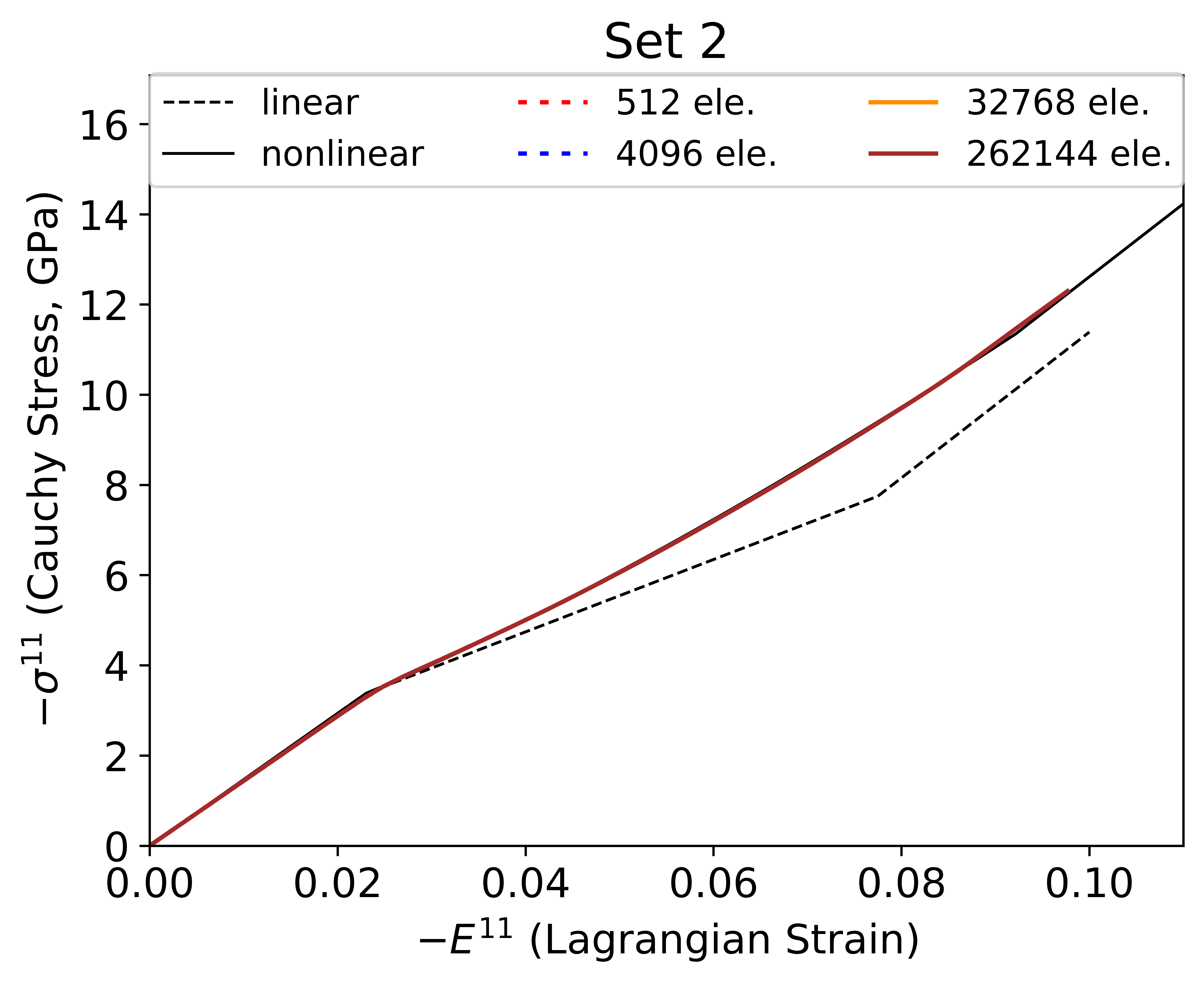}}}\hfill
                \subfloat[\label{3pbc_x_py}]{\resizebox{70mm}{!}{\includegraphics[width=0.5\textwidth]{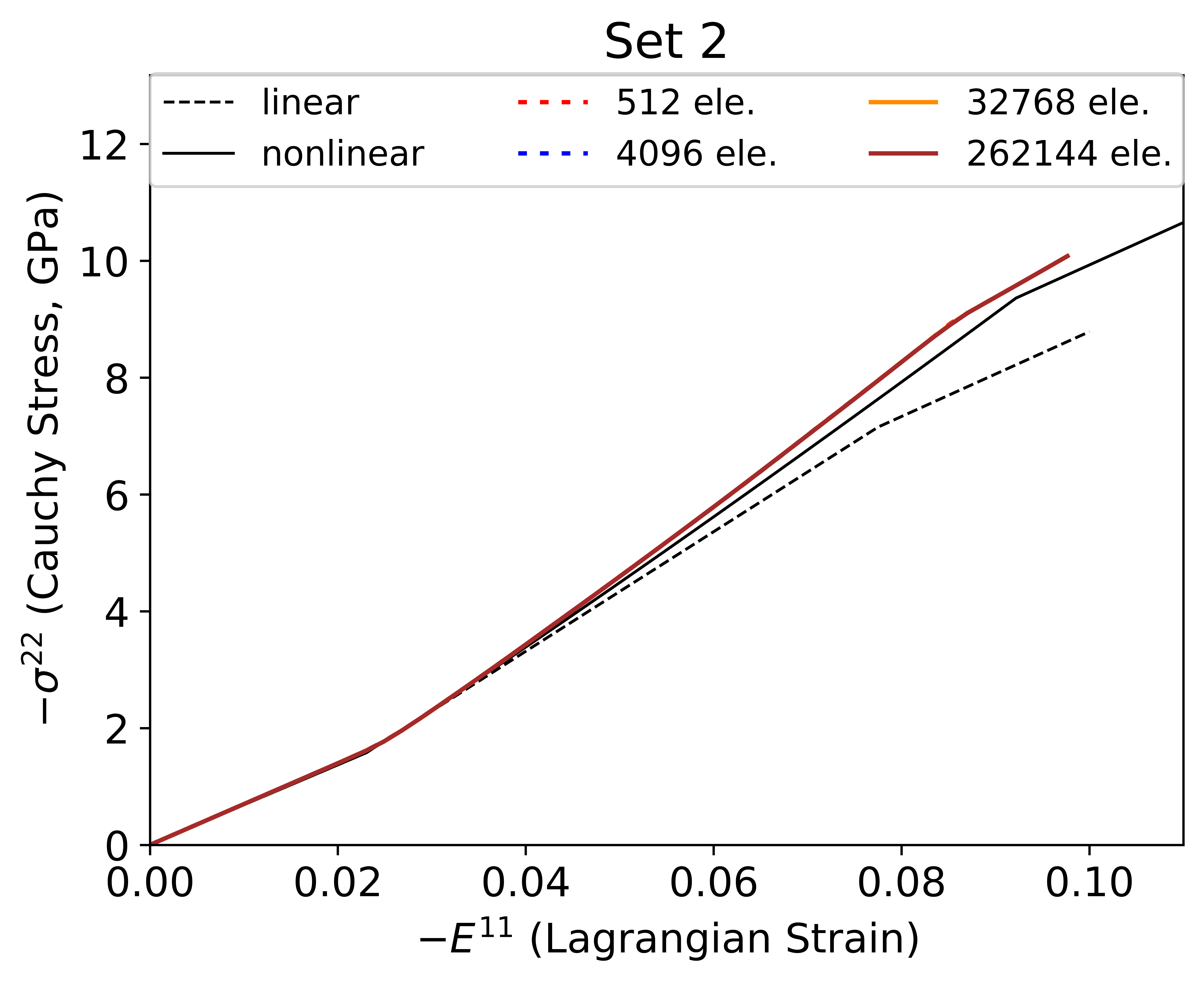}}}\hfill
                \subfloat[\label{3pbc_x_pz}]{\resizebox{70mm}{!}{\includegraphics[width=0.5\textwidth]{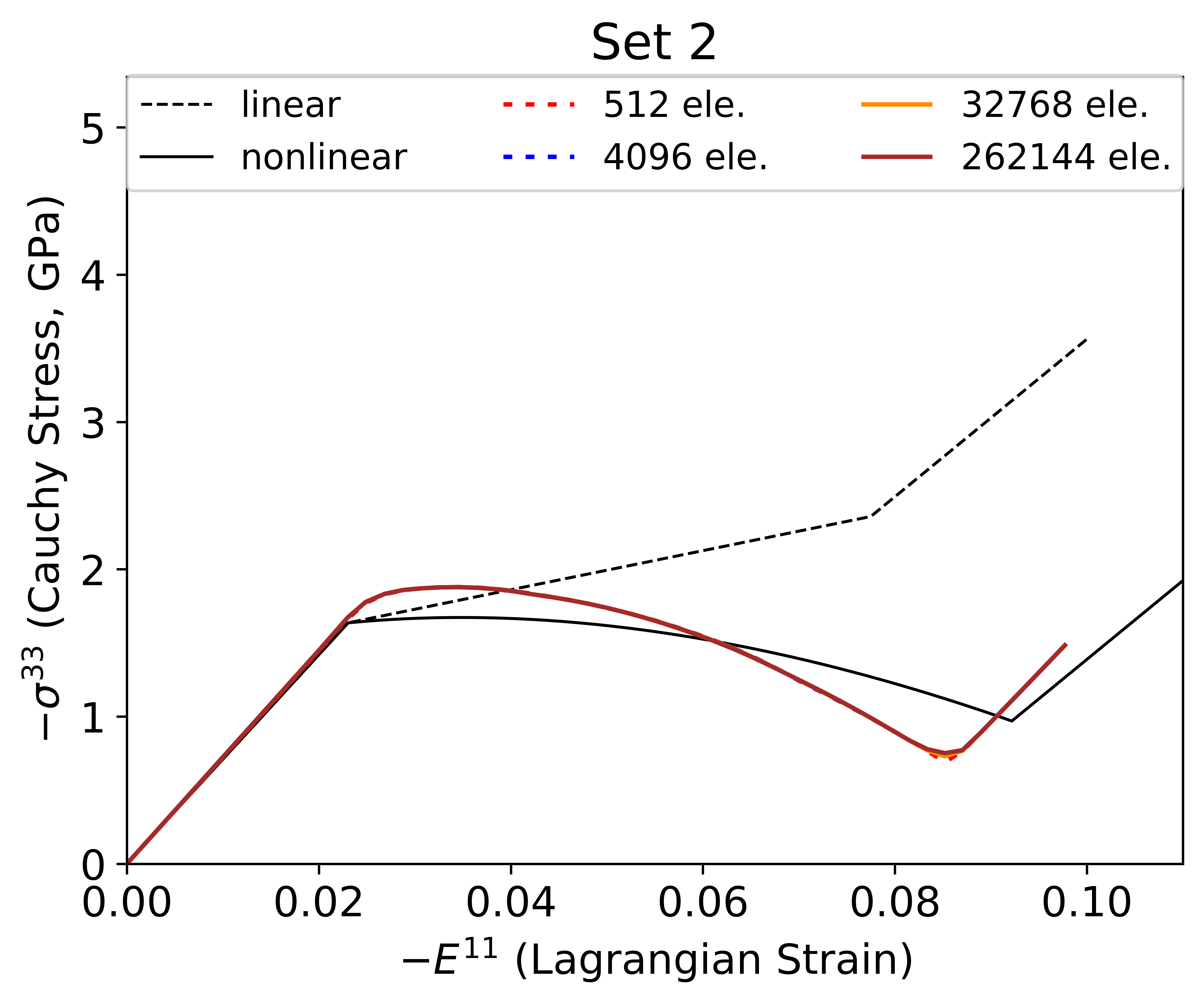}}}
                \caption{Comparison between the analytical and finite element calculations with different number of elements for averaged Cauchy stress $\sg_{ii}$ - Lagrangian strain $E_{11}$ plot for single crystal Zr subjected to the periodic boundary conditions in all three directions and compressive strain in direction 1.}
                \label{3pbc_x}
            \end{figure}

            \Cref{3pbc_x_px,3pbc_x_py,3pbc_x_pz} show the stress-strain plots of different number of elements along with the analytical plot for $\sigma^{11}, \sigma^{22}, \sigma^{33}$ vs $E^{11}$.
            {Analytically, the PT initiates at the principle stresses~$\{-3.380, -1.578, -1.636\}$ GPa and a strain of $-0.023$ and these match exactly with that of simulations.
            The PT ends at the principle stresses $\{-7.748, -7.151, -2.358\}$ GPa and a strain of $-0.077$ with the linear approximation and with nonlinearly varying elastic constants at $\{-11.349, -9.360, -0.970\}$ and $-0.092$.
            In the simulations, the ending stresses are $\{-10.676, -9.107, -0.768\}$ GPa and the strain is $-0.087$.
            In \Cref{3pbc_x_px,3pbc_x_py} we notice a close match between the nonlinear analytical and the simulation calculations, whereas there is a significant mismatch in \Cref{3pbc_x_pz}.
            But for the linear analytical solution, the mismatch is significant for all the cases, with \Cref{3pbc_x_pz} showing no instability as opposed to the simulations and the nonlinear analytical solution.
            This is also observed in \Cref{ineq} not being satisfied for  Zr.
            The difference between the two analytical solutions shows the importance of considering the nonlinearity (i.e., $c$-dependence) in the elastic constants for the PT simulations, especially for  Zr.
            The mismatch between the nonlinear calculations and the simulations is due to the small strain formulation being used in analytical calculations.
            This causes a mismatch in the estimation of the stresses and strains, but the qualitative behavior of the PT is captured well.}

            Although it is expected that there should be a relaxation in the stresses and softening during transformation along the loading axis ($\sigma^{11}$) as an indicator of instability,  similar to what was observed in \Cref{Set1}, we only observe a hardening.
            Similar hardening is observed for $\sigma^{22}$, and only for  $\sigma^{33}$, some softening takes place  indicating PT causing slightly heterogeneous microstructure.
            The reason for such an unusual behavior was discussed in the \Cref{analytical}, that the instability condition (\Cref{ineq}) is satisfied only for $\sigma^{33}$.
            At the same time, increased PT stress to obtain significant amount of $\omega$ phase or complete PT is qualitatively consistent with static experiments  ~\cite{Kumaretal-Acta-20,Kumaretal-Scripta-24,Pandey-Levitasetal-JAP-Zr-24}.
           Also, for dynamic loading,  $\alpha\rightarrow\omega$ PT is observed at higher pressure than under hydrostatic loading (e.g., 9 GPa above the phase equilibrium pressure~\cite{Greeffetal-PRB-22})  or is not observed at all since PT to $\beta$ phase occurs earlier at  $<$20 GPa~\cite{Grivickasetal-JAP-22}.

        \subsection{Set 3}\label{Set3}
            For this study, we do not use periodic boundary conditions but symmetric boundary conditions are applied in the three perpendicular directions on the negative faces in order to simulate a quarter of the sample.
            The sample is subjected to compression in direction 1.
            The loading is applied at a strain rate of $2\times 10^{-6} s^{-1}$ in the elastic regime and then reduced to $ 10^{-9} s^{-1}$ just before the PT.

\begin{figure}[htp]
                \centering
                \subfloat{\resizebox{150mm}{!}{\includegraphics[width=1.0\textwidth]{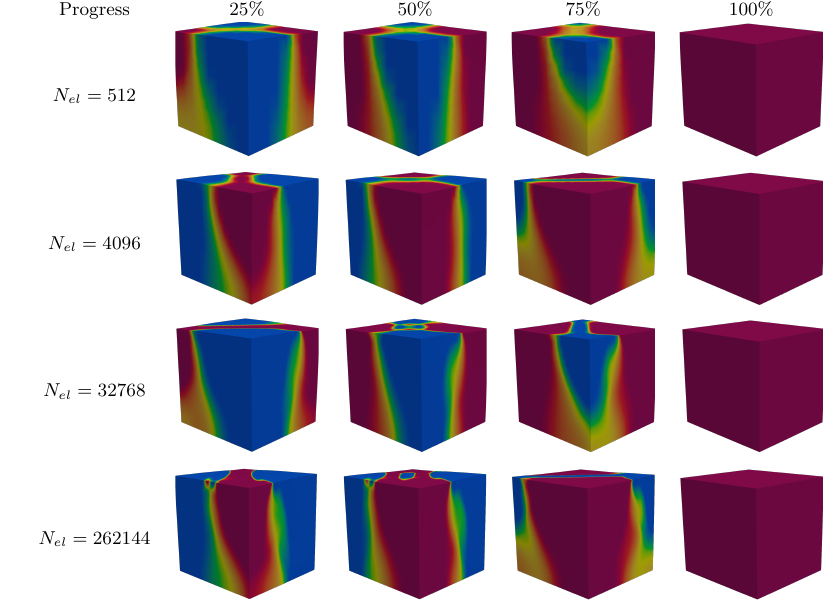}}}\hfill
                \subfloat{\hspace{10mm}\resizebox{20mm}{!}{\includegraphics[width=0.2\textwidth]{Images/axes.png}}}\hspace{35mm}
                \subfloat{\resizebox{60mm}{!}{\includegraphics[width=0.6\textwidth]{Images/colorbar_c.png}}}\hspace{30mm}
                \caption{Evolution of $\omega$ phase in single crystal Zr under symmetric boundary conditions applied on the negative faces of all the three directions and compressive strain in direction 1 shown only for the simulated part without the mirrored parts and varying number of elements.
                The $\omega$ phase shown is the total martensite and consists of variant 1 only.
                The numbers at the top indicate different stages of completion of the simulation for different number of elements.}
                \label{sym_x_c}
            \end{figure}

\begin{figure}[htp]
                \centering
                \subfloat{\resizebox{150mm}{!}{\includegraphics[width=1.0\textwidth]{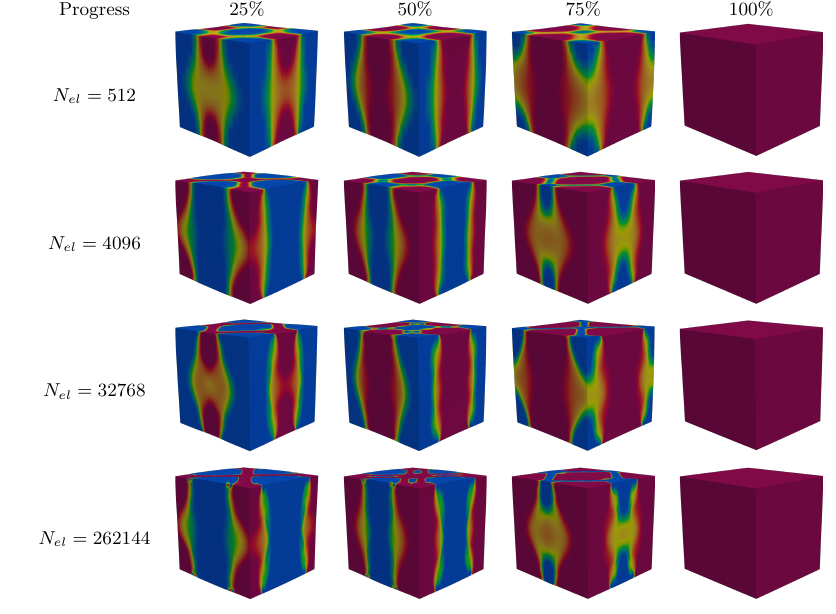}}}\hfill
                \subfloat{\hspace{10mm}\resizebox{20mm}{!}{\includegraphics[width=0.2\textwidth]{Images/axes.png}}}\hspace{35mm}
                \subfloat{\resizebox{60mm}{!}{\includegraphics[width=0.6\textwidth]{Images/colorbar_c.png}}}\hspace{30mm}
                \caption{The same solution like in \Cref{sym_x_c} but shown for the complete sample with mirrored parts and varying number of elements.
                The numbers at the top indicate different stages of completion of the simulation for different number of elements.}
                \label{sym_x_cf}
            \end{figure}

\begin{figure}[htbp]
                \centering
                \resizebox{70mm}{!}{\includegraphics{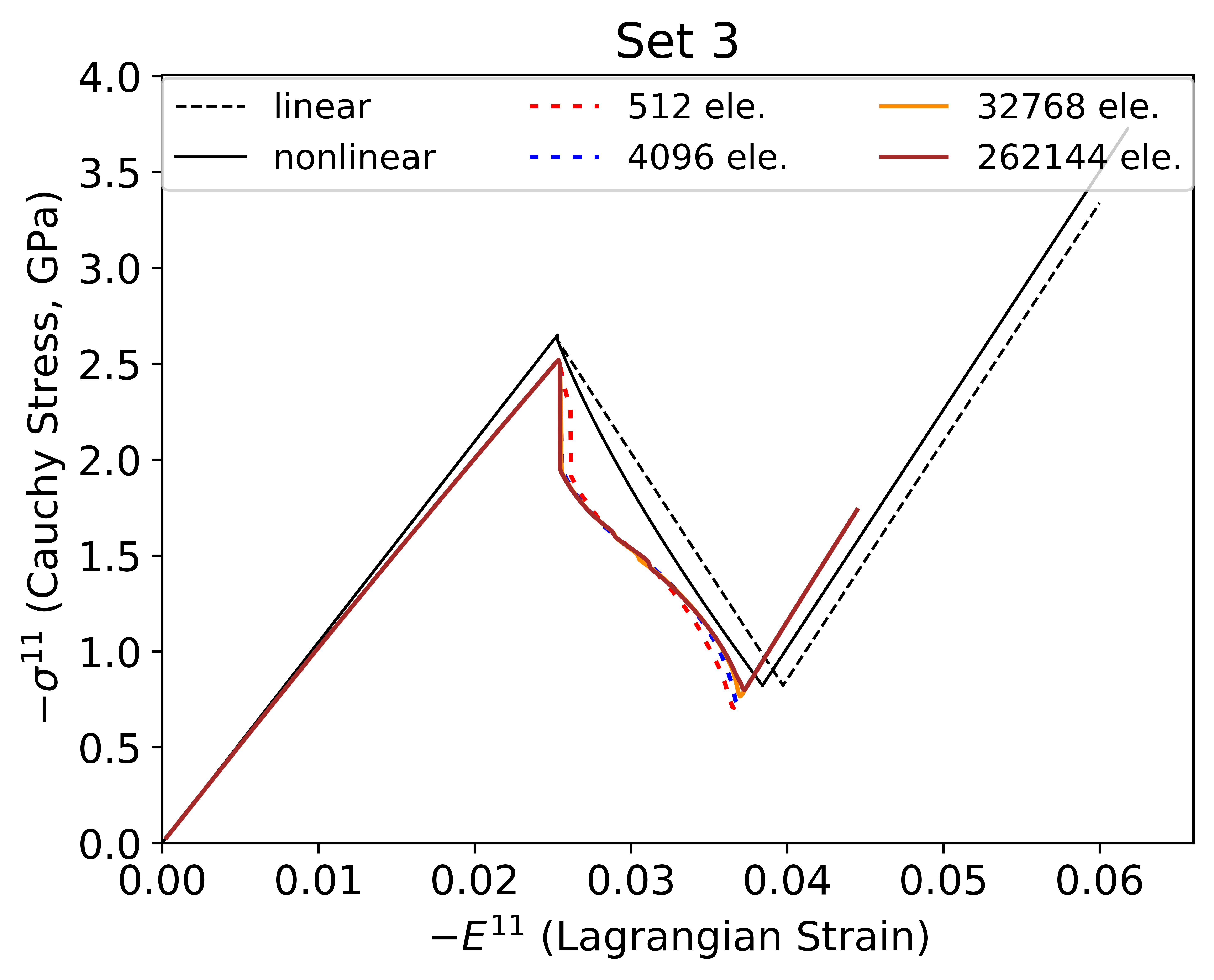}}
                \caption{Comparison between the analytical solution and finite element calculations with different number of elements for averaged Cauchy stress - Lagrangian strain plot for single crystal Zr  with symmetric boundary conditions applied on the negative faces of all the three directions and compressive strain in direction 1}
                \label{sym_x_p}
            \end{figure}

            Like in \Cref{Set1,Set2}, only the first variant appears with this set of boundary conditions. \Cref{sym_x_c} shows the simulated quarter of a sample with the progressive evolution of the first variant and \Cref{sym_x_cf} demonstrates the same in the complete sample including the mirror images in all the three directions.
            Unlike the above cases, even with 512 elements we notice inhomogeneity, the consequence of removing periodic boundary conditions.
            The microstructure developed for 512 and 32,768 elements is similar and likewise solutions for 4,096 and 262,144 elements show similar microstructure. \Cref{sym_x_p} exhibits the stress-strain plots for different number of elements along with the analytical plot.
            {Analytically, PT initiates at a stress of $-2.650$ GPa and a strain of $-0.025$ while in simulations it occurs at a stress of $-2.520$ GPa at the same strain.
            The difference is due to small-strain analytical solution and finite-strain FEM results.
            The PT ends at a stress of $-0.822$ GPa and a strain of $-0.040$ for linear model and a strain of $-0.038$ for the nonlinear calculation.
            For the simulations, the ending stress is $-0.770$ GPa and the strain is $-0.037$.
            This plot shows a deviation in the analytical and simulation plots during the phase transformation  due to the inhomogeneity in the microstructure, as observed in \Cref{sym_x_c,sym_x_cf}.
            However, all FEM results for different number of elements are very close.}

            The difference in the geometric features of the martensitic microstructure for different number of elements is not surprising and acceptable.
            Due to uniform material and stress-strain state, place where material instability starts is determined by small perturbations, namely by error in the numerical procedure during the iterative process.
            It is clear that these places may be different for different number of elements.
            However, practical coincidence of the microstructure for 512 and 3,2768 elements as well as for 4,096 and 262,144 elements confirms that the difference is not directly related to the number of elements.
            All microstructures have very similar vertical and horizontal features, and coincidence of the simulated stress-strain (and, consequently, volume fraction of martensite - strain) curves means that solutions with different number of the finite element are statistically equivalent.
            In experiments for single crystals, microstructure is also repeatable in the statistical sense only.

            Note that the plate-like martensite morphology obtained above is observed in experiments ~\cite{Song-Gray-PMA-95,tewari2008microstructural,banerjee2022omega}.

        \subsection{Set 4}\label{Set4}
            Similar to \Cref{Set2}, periodic boundary conditions are utilized in all three directions.
            In addition, sample is subjected to a compressive strain along direction 3.
            The strain rate of $10^{-5} s^{-1}$ in the elastic regime is applied and then reduced to $ 10^{-11} s^{-1}$ just before the transformation. 

\begin{figure}[htp]
                \centering
                \subfloat{\resizebox{150mm}{!}{\includegraphics[width=1.0\textwidth]{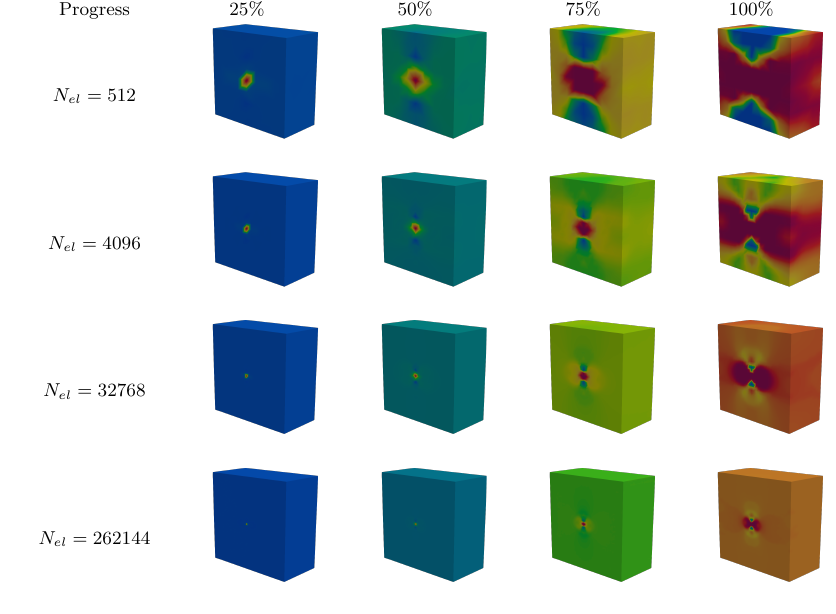}}}\hfill
                \subfloat{\hspace{10mm}\resizebox{20mm}{!}{\includegraphics[width=0.2\textwidth]{Images/axes.png}}}\hspace{35mm}
                \subfloat{\resizebox{60mm}{!}{\includegraphics[width=0.6\textwidth]{Images/colorbar_c.png}}}\hspace{30mm}
                \caption{Evolution of $\omega$ phase in single crystal Zr containing an initial nucleus of variant 1 subjected to  periodic boundary conditions in all three directions and compressed in direction 3 for varying number of finite elements.
                The sample is cut in half along direction 1 to show the internal view.
                The $\omega$ phase shown is the total martensite and has contributions from all the three variants.
                The numbers at the top indicate different stages of completion of the simulation for different number of elements.}
                \label{3pbc_z_c}
            \end{figure}

\begin{figure}[htp]
                \centering
                \subfloat{\resizebox{150mm}{!}{\includegraphics[width=1.0\textwidth]{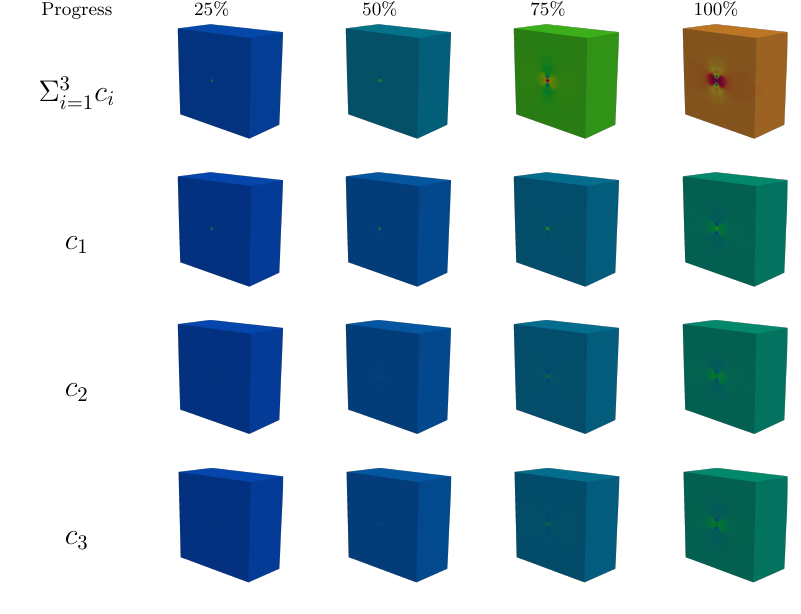}}}\hfill
                \subfloat{\hspace{10mm}\resizebox{20mm}{!}{\includegraphics[width=0.2\textwidth]{Images/axes.png}}}\hspace{35mm}
                \subfloat{\resizebox{60mm}{!}{\includegraphics[width=0.6\textwidth]{Images/colorbar_c.png}}}\hspace{30mm}
                \caption{Evolution of total $\omega$ phase and the individual variants in single crystal Zr containing an initial nucleus of variant 1 subjected to the  periodic boundary conditions in all three directions and compressed in direction 3 for 262,144 elements.
                The sample is cut in half along direction 1 to show the internal view.
                The numbers at the top indicate different stages of completion of the simulation for different number of elements.}
                \label{3pbc_z_ci}
            \end{figure}

            From the transformation strains given in \Cref{epst}, it can be seen that while compressing in direction 3, all the three variants have the same transformation strain.
            Hence, there should be equal volume fractions of all the three variants.
            This was observed in simulations with all the different number of elements.
            In order to observe a distinct microstructure, an initial spherical nucleus of the first variant is introduced with a sphere passing through corners of the cubic FE, as shown in \Cref{3pbc_z_c}.
            For the half of the nucleus radius, $c_1=1$; then it gradually reduces to $c_1=0$ at the spherical boundary. \Cref{3pbc_z_c} shows the half-cut view of the sample and the evolution of the total volume fraction of the martensite for different number of elements. \Cref{3pbc_z_ci} shows the evolution of the total martensite and all the three individual variants for 262,144 elements.
            The nucleus initially partially transforms into second and third variants faster than the rest of the sample is  transforming into martensite.
            But the variant 1 remains  dominant in the nucleus  during the entire simulation.
            The reverse PT in the nucleus does not occur while is not prohibited.
             
\begin{figure}[htbp]
                \centering
                \resizebox{70mm}{!}{\includegraphics{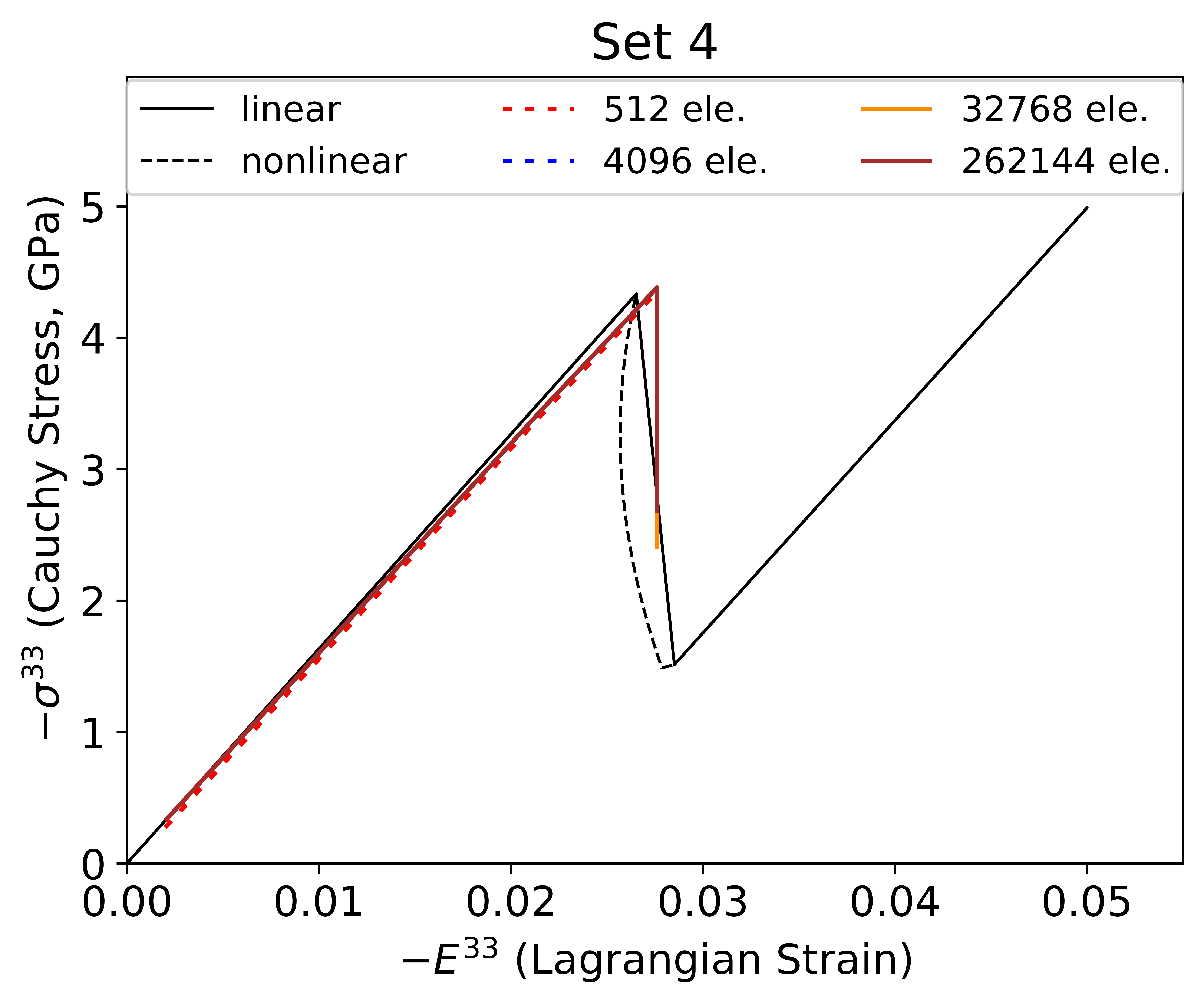}}
                \caption{Comparison between the analytical and finite element calculations with different number of elements for averaged Cauchy stress - Lagrangian strain plot for single crystal Zr containing an initial nucleus of variant 1 under periodic boundary conditions in all three directions and compression in direction 3.}
                \label{3pbc_z_p}
            \end{figure}

             \Cref{3pbc_z_p} shows the stress-strain plot for different number of elements along with the analytical plot.
            During the phase transformation, the linear analytical plot shows an instability with a sharp  with negative slop;
            in nonlinear model,  decrease in both stress and strain occurs, initially with a positive slope which gradually 
            turns negative.
            Since the simulations are strain-controlled, it impossible to follow the analytical strain loading path to accurately replicate it.
            Hence, the strain is held constant after the phase transformation begins.
            With this loading, the stress drastically reduces,  which causes the simulation to diverge before completing the phase transformation in the entire sample.
            Difference in solutions for different number of element is related to reduced size of nucleus with increasing number of FE.
            Complete transformation occurs near nucleus for large number of FE and spreads over the sample for smaller number of FE.
            Analytically, PT initiates at a stress of $-4.333$ GPa and a strain of $-0.026$ while in simulations it occurs at a stress of $-4.381$ GPa and $-0.028$ strain.
            The difference is due to small-strain analytical solution and finite-strain FEM results.
            The PT ends at a stress of $-1.513$ GPa and a strain of $-0.028$ for linear approach and a strain of $-0.028$ and a stress of $-1.488$ GPa for the nonlinear calculation.
            
        \subsection{Set 5}\label{Set5}
            For this study, hydrostatic stress is applied along with symmetry boundary conditions.
            The stress is applied on the positive faces, while symmetry is applied on the negative faces of the three perpendicular directions.
            During the elastic regime, the stress is applied at a rate of $5\times 10^{-3} GPa/s$ and at $2.5\times 10^{-5} GPa/s$ once the phase transformation starts.

\begin{figure}[htp]
                \centering
                \subfloat{\resizebox{150mm}{!}{\includegraphics[width=1.0\textwidth]{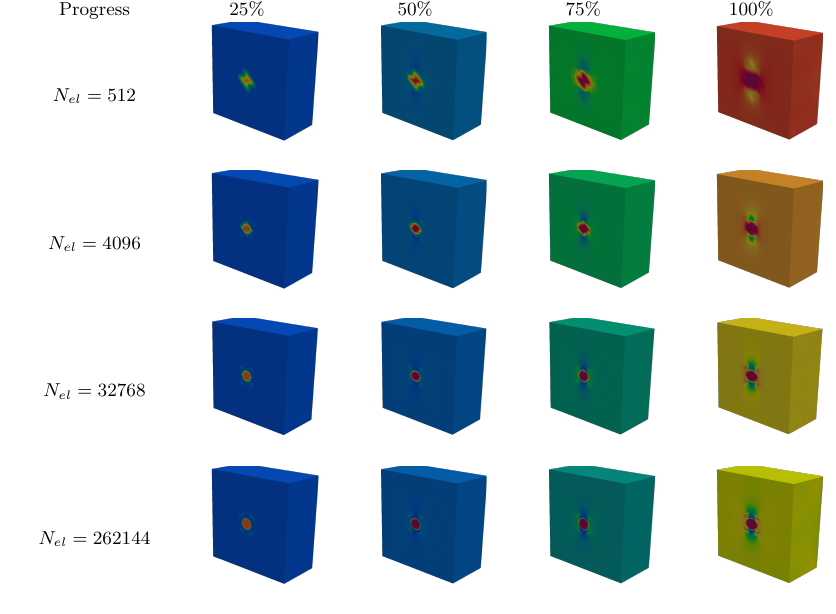}}}\hfill
                \subfloat{\hspace{10mm}\resizebox{20mm}{!}{\includegraphics[width=0.2\textwidth]{Images/axes.png}}}\hspace{35mm}
                \subfloat{\resizebox{60mm}{!}{\includegraphics[width=0.6\textwidth]{Images/colorbar_c.png}}}\hspace{30mm}
                \caption{Evolution of $\omega$ phase in single crystal Zr containing an initial nucleus of variant 1 subjected to hydrostatic loading with varying number of elements.
                The sample is cut in half along direction 1 to show the internal view.
                The $\omega$ phase shown is the total martensite and has contributions from all the three variants.
                The numbers at the top indicate different stages of completion of the simulation for different number of elements.}
                \label{hydro_c}
            \end{figure}

\begin{figure}[htbp]
                \centering
                \resizebox{70mm}{!}{\includegraphics{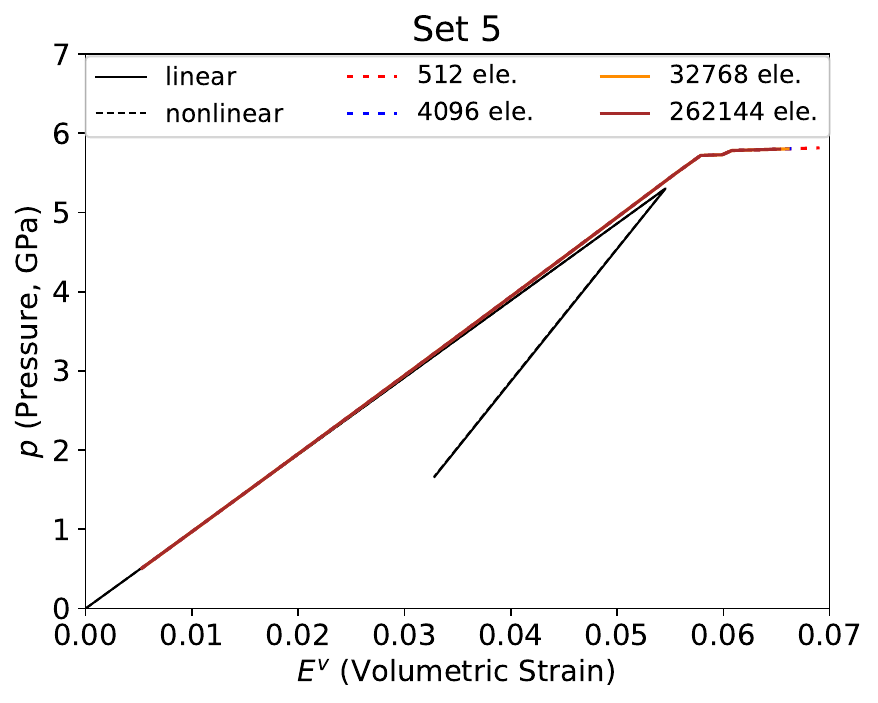}}
                \caption{Comparison between the finite element calculations with different number of elements for averaged Cauchy stress - Lagrangian strain plot for single crystal Zr containing an initial nucleus of variant 1 subjected to hydrostatic loading.}
                \label{hydro_p}
            \end{figure}

            Similar to \Cref{Set4}, the hydrostatic compression resulted in homogeneous and equal volume fractions of all the three variants.
            So, we again introduce a nucleus of the first variant at the center of a sample with the same distribution of $c_1$  as in \Cref{Set4} but with the radius equal to $1/8$ of the cube size for all number of elements. \Cref{hydro_c} shows the evolution of the total martensite for different number of elements.
            Phase transformation is not complete because  the simulation  diverges after $\sim50\%$ of  transformation. \Cref{hydro_p} shows the stress-strain curve for this case.
            Here, we use the pressure and the total volumetric strain instead of the axial stresses and strains.
         {Analytically, the PT starts at a pressure of $-5.31$ GPa and a volumetric strain of $-0.054$ while in simulations it occurs at a volumetric strain of $-0.058$ and a pressure of $-5.72$ GPa.
           The analytical results show a positive elastic modulus as a result of \Cref{ineq5} not being satisfied.
           However, both stress and strain reduces, exhibiting unstable behavior, which is probably the main reason for the computational divergence at prescribes pressure. 
            Unlike the other sets, the two analytical solutions exactly match.
            This is because the bulk compliance is very close for both austenite and martensite in Zr.}

\begin{figure}[htp]
                \centering
                \subfloat{\resizebox{150mm}{!}{\includegraphics[width=1.0\textwidth]{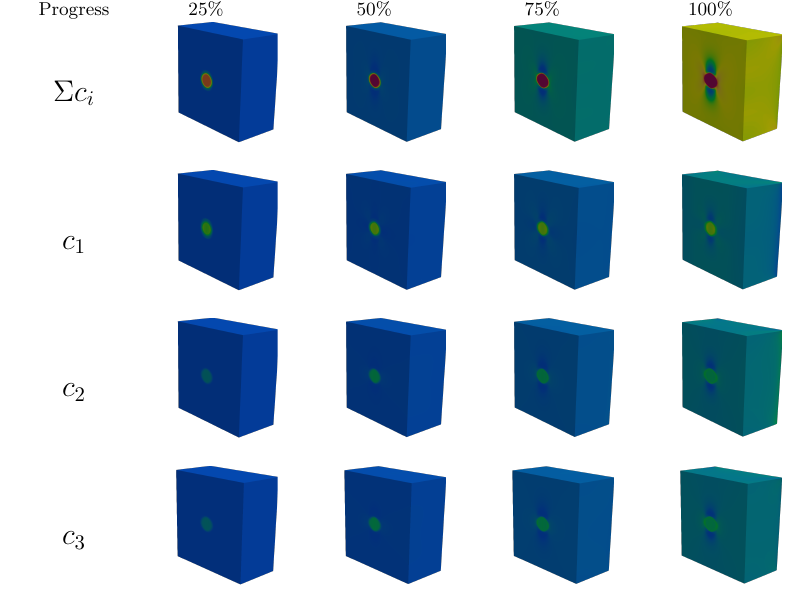}}}\hfill
                \subfloat{\hspace{10mm}\resizebox{20mm}{!}{\includegraphics[width=0.2\textwidth]{Images/axes.png}}}\hspace{35mm}
                \subfloat{\resizebox{60mm}{!}{\includegraphics[width=0.6\textwidth]{Images/colorbar_c.png}}}\hspace{30mm}
                \caption{Evolution of total $\omega$ phase and the individual variants in single crystal Zr containing an initial nucleus of variant 1 subjected to hydrostatic loading for 262,144 elements.
                The sample is cut in half along direction 1 to show the internal view.
                The numbers at the top indicate different stages of completion of the simulation for different number of elements.}
                \label{hydro_6c}
            \end{figure}

            Difference in the volume fraction of martensite in  \Cref{hydro_c} for different number of FE is not directly related to the mesh-dependence of solution but rather to different final PT progress before numerical divergence.
            The large element size, the later divergence starts.
            Within nucleus, variant 1 transforms equally to variants 2 and 3, but still slightly dominates at the beginning of divergence (\Cref{hydro_6c}).
            Outside of nucleus, volume fractions of different variants are approximately the same.

        \subsection{Set 6}\label{Set6}
            For this study, a polycrystal sample with 30 grains having random orientations to eliminate the effects of texture is compressed in direction 1.
            Periodic boundary conditions is applied in all the three directions.
            Of the 30 grains, 2 grains are selected to showcase the PT in a complex loading scenario each of the grains experience due to the load transfer from the neighboring grains.
            The selected grains are oriented at different directions, $\{-0.5063, -3.372, -0.1526\}$ and $\{0.5435, -0.2078, -0.7922\}$, in the Rodrigues notation.
            The sample is loaded at a strain rate of $1.6\times 10^{-4}/s$ during the elastic regime and then at rate of $5\times 10^{-8}/s$.

\begin{figure}[htp]
                \centering
                \subfloat{\resizebox{150mm}{!}{\includegraphics[width=1.0\textwidth]{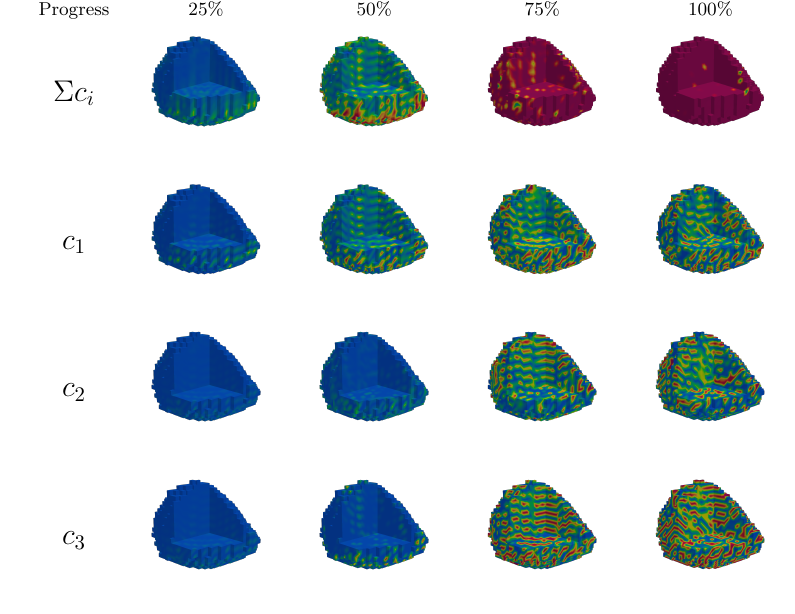}}}\hfill
                \subfloat{\hspace{10mm}\resizebox{20mm}{!}{\includegraphics[width=0.2\textwidth]{Images/axes.png}}}\hspace{35mm}
                \subfloat{\resizebox{60mm}{!}{\includegraphics[width=0.6\textwidth]{Images/colorbar_c.png}}}\hspace{30mm}
                \caption{Evolution of total $\omega$ and different variants in single crystal selected from a polycrystal Zr sample with periodic boundary conditions in all three directions and compressed in direction 1 showing internal and external views.
                The grain shown has the Rodrigues orientation of $\{-0.5063, -3.372, -0.1526\}$ with respect to the global axes.
                The numbers at the top indicate different stages of completion of the simulation for different number of elements.}
                \label{poly_8}
            \end{figure}

\begin{figure}[htp]
                \centering
                \subfloat{\resizebox{150mm}{!}{\includegraphics[width=1.0\textwidth]{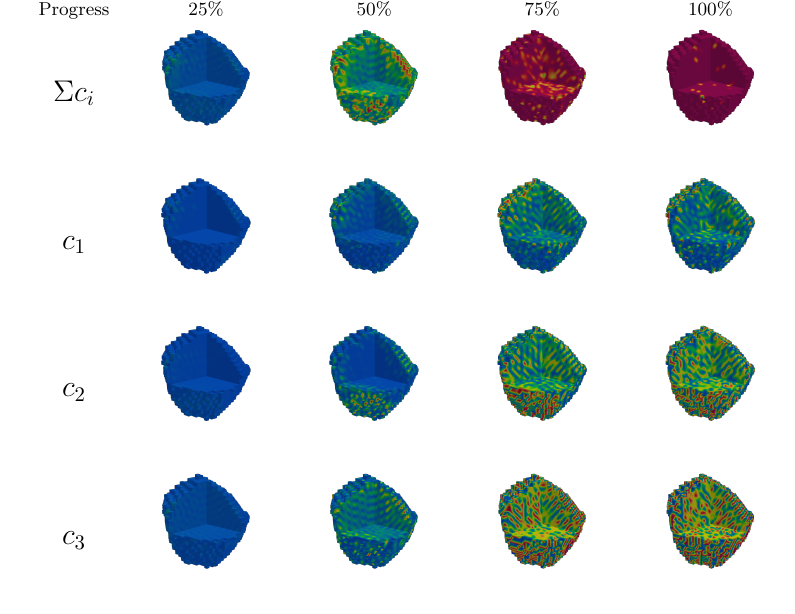}}}\hfill
                \subfloat{\hspace{10mm}\resizebox{20mm}{!}{\includegraphics[width=0.2\textwidth]{Images/axes.png}}}\hspace{35mm}
                \subfloat{\resizebox{60mm}{!}{\includegraphics[width=0.6\textwidth]{Images/colorbar_c.png}}}\hspace{30mm}
                \caption{Evolution of total $\omega$ and different variants in single crystal selected from a polycrystal Zr sample with periodic boundary conditions in all three directions and compressed in direction 1 showing internal and external views.
                The grain shown has the Rodrigues orientation of $\{0.5435, -0.2078, -0.7922\}$ with respect to the global axes.
                The numbers at the top indicate different stages of completion of the simulation for different number of elements.}
                \label{poly_11}
            \end{figure}

            \Cref{poly_8,poly_11} show both the internal and external view of the two selected grains during the different stages of phase transformation of the total martensite as well as the individual martensitic variants.
            The internal view is generated by removing an eighth from the total grain.
            We can notice the difference in the microstructures due to the difference in orientation of the grains.
            In \Cref{Set1,Set2,Set3}, only the first variant appeared when loaded in direction 1.
            But here,  misorientation of direction 1 and the global loading direction and the complex internal stresses from the neighboring grains give rise to a combination of all the three variants and fine martensitic microstructure.
            Specifically, in the selected grains, the variant 3 is more dominant than the others.
            At different transformation stages, martensite morphology includes plate-like and more equiaxial units, as observed in experiments ~\cite{Song-Gray-PMA-95,tewari2008microstructural,banerjee2022omega}.
            Also, the entire $\alpha$ grain transforms into $\omega$ grain, like in TEM observations ~\cite{tewari2008microstructural,banerjee2022omega}, and in situ Laue diffraction results ~\cite{popov2019real,Levitasetal-grainGrowth-24}.

    \section{Some comparisons and interpretations}\label{Comparisons}

        The only existing PFA simulations for $\alpha-\omega$ PT in Zr is performed under hydrostatic loading ~\cite{yeddu2016alpha,Yeddu-Zr-22}, in which various interesting and important results are obtained.
        It shows very fine multivariant martensitic microstructure,  which is not reported in experimental ~\cite{Song-Gray-PMA-95,tewari2008microstructural,banerjee2022omega} and
        molecular dynamic ~\cite{Zong-etal-NPJ-18,Zong-etal-ActaMat-19} works.
        One of the main reasons is that very low interfacial energy of the $\alpha-\omega$ phase interface, $\gamma_{\alpha\omega}=0.01 J/m^2$, was used in simulations.
        Since PFA in ~\cite{yeddu2016alpha,Yeddu-Zr-22} does not have a separate energy barrier and the gradient energy coefficient for martensitic variants, the variant-variant interfacial energy has the same order of magnitude.

        Below, we estimate the energy $\gamma_{\alpha\omega}$ utilising Eq. (74) from the PFA solution in~\cite{Levitas2003}:
        \begin{equation}
        \gamma_{\alpha\omega}=\frac{s_1}{3p_4}\delta_{\alpha\omega}={2.0 G}_{max}\delta_{\alpha\omega}.
        \end{equation}
        Here $\delta_{\alpha\omega}$ is the width of the $\alpha-\omega$ phase interface and $G_{max}$ is the energy barrier between $\alpha$ and $\omega$ phases at the phase equilibrium,  $s_1=16{\ G}_{max}$, and $p_4=2.667$.
        Note that in ~\cite{Yeddu2012},  a similar expression $\gamma_{\alpha\omega}={1.9\ G}_{max}\delta_{\alpha\omega}$ was obtained with a slightly different definition of the $\delta_{\alpha\omega}$.
        The first principle simulations in~\cite{Gao2016} produce  ${G}_{max}=0.02 eV/atom=1.9297  kJ/mol$.
        Assuming $\delta_{\alpha\omega}=0.5\ nm$, we obtain $\gamma_{\alpha\omega}=0.138\ J/m^2$, i.e., more than an order of magnitude larger  than the generic $\gamma_{\alpha\omega}=0.01 J/m^2$ used in ~\cite{yeddu2016alpha,Yeddu-Zr-22}.
        If $\delta_{\alpha\omega}=1\ nm$ is taken (like in ~\cite{Yeddu2012,yeddu2016alpha,Yeddu-Zr-22})
        $\gamma_{\alpha\omega}$ increases to $=0.276 J/m^2$.
        With   such large interfacial energy, much different and less refined microstructure is expected.

        Our scale-free  PFA is not designed to separate martensitic variants, but still some clear separations are observed for the grains of the polycrystal
        (\Cref{poly_8,poly_11}).
        In these pictures, martensite morphology includes plate-like and more equiaxial
        units, as observed in experiments ~\cite{Song-Gray-PMA-95,tewari2008microstructural,banerjee2022omega}.
        Also, the entire $\alpha$ grain transforms into $\omega$ grain,
        like in TEM observations ~\cite{tewari2008microstructural,banerjee2022omega}, and in situ Laue diffraction results ~\cite{popov2019real,Levitasetal-grainGrowth-24}.
        For a single crystals under different boundary conditions, plate-like microstructure also resembles that in experiments ~\cite{Song-Gray-PMA-95,tewari2008microstructural,banerjee2022omega}.

        Surprisingly, for very high yield strength, i.e., without plastic accommodation, PT does not start in ~\cite{yeddu2016alpha}.
        Here, the $\alpha-\omega$ PT occurs without plastic deformations. 
        It is obtained in ~\cite{yeddu2016alpha} that both direct and reverse PTs occur at the same pressure of 4 GPa, which they wrote is in accordance with
        experimental results in~\cite{Zilbershtein-75} were both direct and reverse transformations occur at 2.2 GPa.
        However, this comparison is not correct.
        In~\cite{Zilbershtein-75} both PTs occur during severe plastic deformation under high-pressure torsion, i.e., these are plastic strain-induced PTs rather than pressure-induced PT.
        These experiments were explained in~\cite{Levitas-PRB-2004} based on  the developed kinetics for strain-induced PTs.
        Equality for pressures for direct and reverse PT does not mean that this pressure is the phase equilibrium pressure, as it was claimed in~\cite{Zilbershtein-75,Blank2013}.
        It was shown in~\cite{Levitas-PRB-2004} that such equality may occur in a broad pressure range provided that the stationary volume fraction of the high-pressure phase is reached.
        Zero hystereses in the PFA solution in ~\cite{yeddu2016alpha} is typical for most solutions with homogeneous boundary conditions, because traditional PFA does not include athermal threshold $k$ and interface motion in both directions starts at the infinitesimal thermodynamic driving force (see~\cite{Levitas-Lee-PRL-07,Levitasetal-IJP-10}, where some methods to introduce an athermal threshold in the PFA were also suggested).

        As it is follows from the analysis after \Cref{hystereses}, change  in shear stress from 0.465 to -0.465 GPa reduces the hysteresis in the thermodynamic driving force and normal stresses to zero for a stress-induced PT.
        This also can contribute to zero hysteresis in experiments   in~\cite{Zilbershtein-75,Blank2013}.

        Based on the critical shear stresses for $\alpha$ Zr  the reduction
        in PT pressure due to deviatoric stress in a single crystal does not exceed 0.65 GPa. 
        For a severely deformed polycrystal, which possesses much higher yield strength, the reduction in the PT pressure is 1.52 GPa.
        This value is not sufficient to describe the reduction in PT pressure
        from 6.0 to 0.67 GPa observed during plastic compression in~\cite{Linetal-MRL-23,Lin-Levitas-ResSquare-22} and even from 5.4 to 2.7 GPa in~\cite{Levitas-etal-NatCom-23}.
        This implies that such  reduction is mostly due to the plastic strain-induced PT mechanisms,
        i.e., nucleation at the tip  dislocation pileups or twins ~\cite{Levitas-PRB-2004,levitas2018high,Levitas_2019,Levitas-IJP-21,Levitas-MT-23}.

    \section{Concluding remarks}\label{Conclusion}

        The finite strain scale-free PFA developed in~\cite{babaei2020finite} for cubic to tetragonal PT in single crystal silicon was modified to study  $\alpha-\omega$ (hcp-simple hexagonal) phase transformations in single crystal zirconium.
        All material parameters have been calibrated based on experimental data from the literature.
        The computational procedures and numerical algorithms are implemented using the deal.II FEM program library.

        The effect of the mesh size on the evolution of microstructure was studied in detail, which is crucial for a scale-free models with material instabilities.
        For very coarse meshes, mostly homogeneous solutions were observed.
        As the mesh became finer, discrete evolving  martensitic microstructures appear.
        Analysis shows that further mesh refinement leads to different but statistically equivalent microstructures, i.e., solutions are getting practically mesh independent.
        For some loadings, e.g., with the symmetry boundary conditions, discrete martensitic microstructure appears even for the roughest mesh.
        For periodic boundary conditions along all sides, analytical solution predicts no material instability in the loading and one lateral directions but possible instability in another lateral direction.
        In the PFA simulations, solutions
        for all meshes are almost homogeneous, with minor band structures at later transformation stage, practically the same for meshes with 32,768 and 262,144 elements.
        Since for loadings in direction 3, for which transformation strains for all variants are the same, and for hydrostatic loading, all variants appear in the same proportions and homogeneously, nucleus of the variant 1 at the center of the sample was introduced to trigger heterogeneous microstructure.

        Much smaller transformation strains of different signs (with maximum compressive strain of -0.0318 and tensile strain of 0.038) with the same third normal components of -0.022 for all three variants lead to completely different microstructure than in Si, which has very large transformation strain components (but without shear).
        In particular, for various simple straining in direction 1 with maximum compressive transformation strain and different conditions at other two surfaces,
        variant 1 appears only, in contrast to complex multivariant microstructure for Si.
        For loadings in direction 3, for which transformation strains for ll variants are the same, and for hydrostatic loading, all variants appear in the same proportions.
        Such relatively simple microstructures allowed us to predict stress-strain and
        $c_i$-strain curves
        for these cases analytically, with good correspondence with PFA simulations.
        The main reason for small discrepancy between analytical and numerical results is formation of discrete  martensitic microstructure, while analytical solution is for the uniform stresses and strains.
        For periodic boundary conditions in all three directions, for which numerical solution is also close to the homogeneous, essential  difference is related to small-strain analytical solution as compared to the finite-strain numerical solutions.
        Thus, even for relatively small transformation and elastic strains,
        finite-strain formulation is important.
        Also, since the transformation strains for Zr were much smaller compared to Si, the strain rates to mimic a quasi-static loading condition are much lower.

        Analytical criteria for initiation of the direct and reverse $\alpha$ to $\omega$ PTs and variant-variant transformations under general stress tensor
        have been derived and analysed in detail.
        Due to small volumetric transformation strain (-0.0153) and larger deviatoric uniaxial transformation strain
        $e_{t1}^{11}=0.0371$ and shear strain of 0.0604, the effect of the deviatoric stresses is strong and should be taken into account not only for analysis of the experiments on triaxial compression in different deformation high-pressure apparatuses, but also for the description of the
        plastic strain-induced PTs.
        The kinetics of strain-induced PTs in general~\cite{Levitas-PRB-2004}, and in Zr in particular
       ~\cite{Pandey-Levitas-2020,Levitas-etal-NatCom-23,Dhar-Levitas-NPJ-kinetics-24,Lin-Levitas-ResSquare-22}, explicitly depends on
        pressure and accumulated plastic strain, and deviatoric stresses are included indirectly only.
        However, during plastic compression in DAC or torsion in rotational DAC, pressure and nonhydrostatic stresses in the central part of the sample grow and may exceed the critical values defined in \Cref{A-M1}-\Cref{M3-A}   for stress-induced PT.
        Thus, stress-controlled kinetics should be added to the strain-controlled kinetics, and then both should be calibrated, which is quite challenging.

        For a single crystals subjected to various  boundary conditions, plate-like microstructure mimics   that in known experiments ~\cite{Song-Gray-PMA-95,tewari2008microstructural,banerjee2022omega}.
        For grains of the polycrystal, martensite morphology includes plate-like and more equiaxial
        units, similar to those in experiments ~\cite{Song-Gray-PMA-95,tewari2008microstructural,banerjee2022omega}.
        Also, the entire $\alpha$ grain transforms into $\omega$ grain, without residual $\alpha$ phase,
        similar to the  TEM results ~\cite{tewari2008microstructural,banerjee2022omega}, and in situ Laue diffraction findings ~\cite{popov2019real,Levitasetal-grainGrowth-24}.

        Obtained results are the basis for studying behaviour of a polycrystalline Zr sample.
        The first step has been already made.
        The aggregate  with 30 randomly orientated grains have been model mainly to produce complex heterogeneous loadings for two selected grains.
        Complex fine discrete microstructure for all martensitic variants was obtained for both grains, illustrating that relatively simple microstructures for all other cases  are related to simple and uniform loadings of single crystals along the crystallographic directions.
        Much more detailed study with much larger grain numbers and complex loading will be performed and reported elsewhere.

        A similar approach can be applied to other $\alpha$ to $\omega$ PTs, e.g., in titanium and hafnium.
        Furthermore, the PT should be combined with the discrete localized plastic flow (e.g., dislocation pileups, shear bands, and twins)~\cite{Levitas2018Scale-FreeMicrostructure,Esfahani-etal-2020} in order to study plastic strain-induced PTs caused by strong stress concentrators at the tip pileups and bands.
        This is required to explain strong reduction in the PT pressure and describe strain-controlled transformation kinetics observed in experiments~\cite{Zilbershtein-75,Blank2013,Srinivasarao-etal-Zr-11,Pandey-Levitas-2020,Levitas-etal-NatCom-23,Dhar-Levitas-NPJ-kinetics-24,Lin-Levitas-ResSquare-22}.

\noindent {\bf Acknowledgement}
\par The authors acknowledge the support of the US National Science Foundation  (DMR-2246991),  the US Army Research Office (W911NF2420145) and Iowa State University (Murray Harpole Chair in Engineering).
The simulations were performed at  NSF Advanced Cyberinfrastructure Coordination Ecosystem: Services \& Support (ACCESS), allocation TG-MSS170015.

\bibliography{ref_updated}

\end{document}